\documentclass[aps,prb,twocolumn,groupedaddress]{revtex4}

\usepackage{graphicx}% Include figure files
\usepackage{dcolumn}% Align table columns on decimal point
\usepackage{bm}% bold math
\usepackage{epstopdf}
\usepackage{amsmath}
\usepackage{color}
\usepackage{units}

\newcommand{\figref}[1]{\mbox{Fig.\,\ref{#1}}}
\renewcommand{\eqref}[1]{\mbox{Eq.\,(\ref{#1})}}
\newcommand{\tabref}[1]{\mbox{Table~\ref{#1}}}
\newcommand{\secref}[1]{\mbox{Sec.~\ref{#1}}}

\newcommand{\rd}{\ensuremath{\mathrm{d}}}
\newcommand{\id}{\ensuremath{\,\rd}}

\newcommand{\ket}[1]{\left|{#1}\right\rangle}  

\newcommand{\ketbra}[2]{\left|{#1}\rangle \langle{#2}\right|}

\newcommand{\comm}[2]{\left[ #1, #2 \right]}
\newcommand{\lind}[1]{\mathcal{D}\left[#1\right]}

\newcommand{\sm}{\sigma_-}
\renewcommand{\sp}{\sigma_+}
\newcommand{\abs}[1]{\left|#1\right|}
\newcommand{\abssq}[1]{\left| #1 \right|^2}
\newcommand{\nn}{\nonumber}

\newcommand{\be}{\begin{equation}}
\newcommand{\ee}{\end{equation}}
\newcommand{\bea}{\begin{eqnarray}}
\newcommand{\eea}{\end{eqnarray}}

\begin{document}

\title{Quantum Acoustics with Surface Acoustic Waves}

\author{Thomas Aref$^1$}
\author{Per Delsing$^{1}$}\email{per.delsing@chalmers.se}
\author{Maria K. Ekstr\"om$^1$}
\author{Anton Frisk Kockum$^1$}
\author{Martin V. Gustafsson$^{1,2}$}
\author{G\"oran Johansson$^1$}
\author{Peter Leek$^3$}
\author{Einar Magnusson$^3$}
\author{Riccardo Manenti$^3$}
\affiliation{$^1$ Microtechnology and Nanoscience, Chalmers University of Technology, G\"oteborg, Sweden.}
\affiliation{$^2$ Department of Chemistry, Columbia University, New York, NY, USA.}
\affiliation{$^3$ Clarendon Laboratory, Department of Physics, University of Oxford, Oxford, United Kingdom}
%\date{\today}

\begin{abstract}
It has recently been demonstrated that surface acoustic waves (SAWs) can interact with superconducting qubits at the quantum level. SAW resonators in the GHz frequency range have also been found to have low loss at temperatures compatible with superconducting quantum circuits. These advances open up new possibilities to use the phonon degree of freedom to carry quantum information. In this paper, we give a description of the basic SAW components needed to develop quantum circuits, where propagating or localized SAW-phonons are used both to study basic physics and to manipulate quantum information. Using phonons instead of photons offers new possibilities which make these quantum acoustic circuits very interesting. We discuss general considerations for SAW experiments at the quantum level and describe experiments both with SAW resonators and with interaction between SAWs and a qubit. We also discuss several potential future developments.
\end{abstract}

\maketitle

\section{Introduction}
\label{sec:intro}
Quantum optics studies the interaction between light and matter. Systems of electromagnetic waves and atoms can be described by quantum electrodynamics (QED) in great detail and with amazingly accurate agreement between experiment and theory. A large number of experiments have been carried out within the framework of cavity QED, where the electromagnetic field is captured in a 3D cavity and allowed to interact with individual atoms. Cavity QED has been developed both for microwaves interacting with Rydberg atoms \cite{HarocheBook} and for optical radiation interacting with ordinary atoms \cite{MillerJPB2005}. In this paper, we will discuss the acoustic analogue of quantum optics, which we might call ``quantum acoustics'', where acoustic waves are treated at the quantum level and allowed to interact with artificial atoms in the form of superconducting qubits.  

The on-chip version of quantum optics is known as circuit QED and has been studied extensively in recent years. In circuit QED, superconducting cavities are coupled to artificial atoms in the form of superconducting electrical circuits that include Josephson junctions \cite{Wallraff,SchoelkopfGirvin}. The nonlinearity of the Josephson junctions is used to create a nonequidistant energy spectrum for the artificial atoms. By isolating the two lowest levels of this energy spectrum, the artificial atoms can also be used as qubits. The most commonly used circuit is the transmon \cite{Koch}. Often the junction is replaced by a superconducting quantum interference device (SQUID), consisting of two Josephson junctions in parallel. This allows the level splitting of the artificial atom to be tuned \textit{in situ} by a magnetic field so that the atom transition frequency can be tuned relative to the cavity resonance frequency. However, a cavity is not necessary to study quantum optics. The interaction between a transmission line and the artificial atom can be made quite large even without a cavity. Placing one or more atoms in an open transmission line provides a convenient test bed for scattering between microwaves and artificial atoms in one dimension. This subdiscipline of circuit QED is referred to as waveguide QED \cite{Zheng,ZhengPRL,Auffeves,HoiNJP}.

In a number of experiments, systems exploiting the mechanical degree of freedom have been investigated, and in several cases they have reached the quantum limit \cite{Teufel2009,LaHaye2009,Oconnell2010,Pirkkalainen2013}. Typically, systems containing micro-mechanical resonators in the form of beams or drums are cooled to temperatures low enough that the thermal excitations of the mechanical modes are frozen out. This can be done in two different ways: either the frequency is made so high that it suffices with ordinary cooling in a dilution refrigerator, or alternatively, for mechanical resonators with lower frequencies, active cooling mechanisms such as sideband cooling can be employed.

With the development of radar in the mid-20th century, a need arose for advanced processing of radio frequency (RF) and microwave signals. An important class of components created to fill this need is based on surface acoustic waves (SAWs), mechanical ripples that propagate across the face of a solid. When SAWs are used for signal processing, the surface of a microchip is used as the medium of propagation. An electrical RF signal is converted into an acoustic wave, processed acoustically, and then converted back to the electrical domain. The substrate is almost universally piezoelectric, since this provides an efficient way to do the electro-acoustic transduction. The conversion is achieved using periodic metallic structures called interdigital transducers (IDTs), which will be discussed later. 

Since the speed of sound in solids is around five orders of magnitude lower than the speed of light, SAWs allow functions like delay lines and convolvers to be implemented in small packages \cite{datta1986, morgan2007, campbell1998}. The acoustic wavelength is correspondingly short, and thus reflective elements and gratings can readily be fabricated on the surface of propagation by lithography. These features enable interference-based functionality, such as narrow-band filtering, with performance and economy that is unmatched by all-electrical devices. Since the heyday of radar development, SAW-based components have found their way into almost all wireless communication technology, and more recently also into the field of quantum information processing.

The most well-explored function of SAWs in quantum technology has hitherto been to provide a propagating potential landscape in semiconductors, which is used to transport carriers of charge and spin \cite{Barnes2000, Hermelin2011, McNeil2011}. Here we are concerned with an altogether different kind of system. Rather than using SAWs to transport particles, we focus on quantum information encoded directly in the mechanical degree of freedom of SAWs. This use of SAWs extends the prospects of mechanical quantum processing to propagating phonons, which can potentially be used to transport quantum information in the same way as itinerant photons.

Most solid-state quantum devices, such as superconducting qubits, are designed to operate at frequencies in the microwave range ($\sim\unit[5]{GHz}$). These frequencies are high enough that standard cryogenic equipment can bring the thermal mode population to negligible levels, yet low enough that circuits much smaller than the electrical wavelength can be fabricated. When cooled to low temperatures, SAW devices show good performance in this frequency range, where suspended mechanical resonators tend to suffer from high losses. Indeed, recent experiments report Q-values above $10^5$ at millikelvin temperatures for SAWs confined in acoustic cavities \cite{Magnusson2014}. Another recent experiment has shown that it is possible to couple SAW waves to artificial atoms in a way very similar to waveguide QED \cite{Gustafsson2014}. Combining these results opens up the possibility to study what corresponds to cavity QED in the acoustic domain.

%By use of a piezoelectric substrate, SAWs can be generated efficiently from electrical signals and converted back to the electrical domain after propagating acoustically over a long distance on the chip.

The advantageous features of sound compared to light are partly the same in quantum applications as in classical ones. The low speed of sound offers long delay times, which in the case of quantum processing can allow electrical feedback signals to be applied during the time a quantum spends in free propagation. The short wavelength allows the acoustic coupling to a qubit to be tailored, distributed, and enhanced compared with electrical coupling.

In this paper we will discuss the exciting possibilities in this new area of research. First, in \secref{sec:MaterialsFabrication}, we describe the SAW devices and how they are fabricated. Then, \secref{sec:Theory} provides a theoretical background. In \secref{sec:ClassicalTheory}, we describe the classical theory needed to evaluate and design the SAW devices, in \secref{sec:SemiClassicalTheory} we present a semiclassical model for the coupling between a qubit and SAWs, and in \secref{sec:QuantumTheory} we present the quantum theory. In the following section, we describe the SAW resonators and their characterization. The SAW-qubit experiment is discussed in \secref{sec:SAWqbInteraction} and in the last section we give an outlook for interesting future experiments.

\section{Surface acoustic waves, materials and fabrication}
\label{sec:MaterialsFabrication}

There are several different types of SAWs, but here we will use the term to denote Rayleigh waves \cite{rayleigh1885, datta1986, morgan2007}. These propagate elastically on the surface of a solid, extending only about one wavelength into the material. At and above radio frequencies, the wavelength is short enough for the surface of a microchip to be used as the medium of propagation. 

The most important component in ordinary SAW devices is the IDT, which converts an electrical signal to an acoustic signal and vice versa. In its simplest form, the IDT consists of two thin film electrodes on a piezoelectric substrate. The electrodes are made in the form of interdigitated fingers so that an applied AC voltage produces an oscillating strain wave on the surface of the piezoelectric material. This wave is periodic in both space and time and radiates as a SAW away from each finger pair. The periodicity of the fingers $\lambda$ defines the acoustic resonance of the IDT, with the frequency given by $\omega_{IDT}=2 \pi f_{IDT}=2 \pi v_0/\lambda$, where $v_0$ is the SAW propagation speed. When the IDT is driven electrically at $\omega=\omega_{IDT}$, the SAWs from all fingers add constructively which results in the emission of strong acoustic beams across the substrate surface, in the two directions perpendicular to the IDT fingers. The number of finger pairs $N_p$ determines the bandwidth of the IDT, which is approximately given by $f_{IDT}/N_p$.

\subsection{Materials for quantum SAW devices}

The choice of substrate strongly influences the properties of the SAW device. The material must be piezoelectric to couple the mechanical SAW to the electrical excitation of the IDT. Because SAW devices have many commercial applications, a wide search for suitable piezoelectric crystals has been performed. Theoretical and experimental material data are available for many substrates, albeit until recently not at the millikelvin temperatures required for quantum experiments. 
%both for impedance matching and to shunt the SQUID forming a transmon qubit. 
%Examples of substrates of interest include lithium niobate, lithium tantalate, quartz and gallium arsenide. One can also deposit piezoelectric films such as aluminum nitride and zinc oxide instead of using a piezoelectric substrate. 
Piezoelectric materials used for conventional SAW devices \cite{morgan2007} include bulk piezoelectric substrates such as gallium arsenide (GaAs), quartz, lithium niobate ($\textrm{LiNbO}_{3}$), and lithium tantalate ($\textrm{LiTaO}_{3}$), and piezoelectric films such as zinc oxide (ZnO) and aluminum nitride (AlN) deposited onto nonpiezoelectric substrates. For any given cut of a piezoelectric crystal, there are only a few directions where SAWs will propagate in a straight line without curving (an effect known as beam steering). Thus it is common to specify both the cut and the direction of a substrate, which also affect piezoelectric properties. For example, YZ lithium niobate is a Y-cut with propagation in the Z direction. Additional effects that need to be considered are diffraction, bulk wave coupling, other surface wave modes, ease of handling, etc. 

\begin{table*}[t]
\centering
\begin{tabular}{| r | r | r | r | r | r |}\hline
		  & Temp. & $\textrm{LiNbO}_{3}$ & GaAs                      & Quartz & ZnO     \\ \hline
Cut/Orient. &            & Y-Z                              & \{110\}-$<$100$>$ & ST-X    & (0001)  \\ \hline
$\unit[K^2]{[\%]}$ & $\unit[300]{K}$ & 4.8 \cite{morgan2007}& 0.07 \cite{datta1986}& 0.14 \cite{morgan2007}&  \\ \hline
$\unit[C_S]{[pF/cm]}$ & $\unit[300]{K}$ & 4.6 \cite{datta1986}& 1.2 \cite{datta1986}& 0.56 \cite{morgan2007}& 0.98 \cite{Madelung1999} \\ \hline
Diffraction $\gamma$ & $\unit[300]{K}$ & -1.08 \cite{Slobodnik1978}& -0.537* \cite{Slobodnik1978}& 0.378 \cite{Slobodnik1978}& --- \\ \hline
$\unit[\delta]{[ppm/K]}$ & $\unit[300]{K}$ & 4.1\cite{Browder1977} & 5.4 \cite{Smith1975}& 13.7 \cite{hashimoto2000}& 2.6 or 4.5 \cite{Yates1971}\\ \hline
$\unit[\Delta l/l]{[\%]}$ & $\unit[4]{K}$ & 0.08 \cite{White2002}& 0.10 \cite{White2002}& 0.26 \cite{White2002} & 0.05 or 0.09 \cite{White2002}\\ \hline
$\unit[v_{0}]{[m/s]}$ & $\unit[300]{K}$ & 3488 \cite{datta1986}& 2864	 \cite{datta1986}& 3158 \cite{datta1986}& --- \\ \hline
$\unit[v_{e}]{[m/s]}$ & $\unit[300]{K}$ & ---	 & 2883$\pm$1 \cite{Smith1975}& 3138$\pm$1 \cite{Smith1975}& --- \\ \hline
$\unit[v_{e}]{[m/s]}$ & $\unit[10]{mK}$ & --- & 2914.3$\pm$0.8 \cite{Smith1975}& 3134.7$\pm$1.5 \cite{Smith1975}& 2678.2$\pm$0.5 \cite{Smith1975}\\ \hline
$\unit[\alpha_{vis}]{[dB/\mu s]}$ & $\unit[300]{K}$ & 0.88 \cite{Slobodnik1978}& 0.9 \cite{Hunt1986}& 2.6 \cite{morgan2007}& --- \\ \hline
$\unit[\alpha_{vis}]{[dB/\mu s]}$ & $\unit[10]{mK}$ & 0.3 \cite{Slobodnik1978}& $<$0.44 \cite{Smith1975}& $<$0.44 \cite{Smith1975}& $<$0.11\cite{Smith1975} \\ \hline
\end{tabular}
\caption{
Material properties for selected substrates at $\unit[300]{K}$ and $\unit[10]{mK}$. The effective velocity is extracted from the resonant frequency of 1-port SAW resonators with $\unit[100]{nm}$ (GaAs and quartz) and $\unit[30]{nm}$ (ZnO) thick aluminum electrodes. GaAs data is for devices on the \{110\} plane and SAW propagation in the $<$100$>$ direction. ZnO data is for devices on the (0001) plane. Diffraction can be quantified by the derivative of the power flow angle $\gamma$ and is minimized when this parameter approaches $-1$ \cite{morgan2007}. $\delta$ is the thermal expansion coefficient.
\newline
*There seems to be some disagreement between different articles \cite{Hunt1991,Slobodnik1978,deLima2003}.
\label{tab:MaterialProperties}}
\end{table*}

Several material properties affect the suitability for quantum experiments and must be fine-tuned with various trade-offs. The two primary factors that play into the design of IDTs and similar structures are the piezoelectric coupling coefficient, $K^2$, and the effective dielectric constant, $C_S$, which is expressed as the capacitance per unit length of a single finger pair. Other parameters of importance are propagation speed $v_0$, finger overlap $W$, attenuation rate $\alpha$, and coefficients for diffraction, $\gamma$ and thermal expansion $\delta$. Table\,\ref{tab:MaterialProperties} summarizes our current knowledge of these parameters for a selection of substrates.

%ZnO is a relatively common material in SAW devices, but until now there have only been reports of its use as a thin layer piezoelectric transducer on top of a non-piezoelectric substrate such as sapphire, diamond or SiO$_2$/Si \cite{morgan2007, Weber}. Wafer scale bulk ZnO has recently become available \cite{Look1998, Look2001, Avrutin2010}, but even nominally pure material contains dopants that give rise to a non-zero room-temperature electrical conductivity. This conductivity damps SAWs efficiently, impeding the use of bulk ZnO in SAW devices.
%
%Due in part to the development of bulk crystal growth, ZnO is now receiving growing interest as a device material \cite{Avrutin2010, Jagadish2006, Ozgur2005}, in particular due to its promise for optoelectronic applications \cite{Ozgur2005}, and is also of significant interest for quantum information devices, due to the presence of long lifetime spin defect centers in the crystal \cite{George2013}.

The material properties relevant for SAW devices may be substantially altered from known textbook values in the high vacuum and low temperature environment used for quantum devices. For example, the attenuation coefficient for SAW propagation loss is given at room temperature in air by \cite{morgan2007}
\begin{equation}
\alpha=\alpha_{\rm{air}}\left(P\right)\frac{f}{10^{9}}+\alpha_{\rm{vis}}\left(T\right)\left(\frac{f}{10^{9}}\right)^{2}\; \left[\frac{dB}{\mu s}\right],
\end{equation}
where the first term is due to air loading and the second due to viscous damping in the substrate. $\alpha_{\rm{air}}$ can be neglected in high vacuum, and although reliable values are not yet available at low temperature for $\alpha_{\rm{vis}}$, an upper limit can be placed on it from known resonator quality factors.

When an IDT device is cooled to low temperature, the frequency of the device changes for two reasons. First, there is a length contraction that alters the distance between the IDT fingers, and also between reflectors in a civity. Second, the speed of sound changes. The fractional length contraction $\Delta l/l$ when cooling down from room temperature to liquid helium temperatures can be estimated as $\Delta l/l=(l_{RT}-l_{4K})(l_{RT})\approx-190\,\delta$ if the Debye temperature is in the range 250-450\,\rm{K} \cite{White2002}. Several of the materials have somewhat higher Debye temperatures, which should lower the values in Table\,\ref{tab:MaterialProperties} slightly.

%Text about velocity:
In Table \ref{tab:MaterialProperties}, two different propagation velocities are reported: the free velocity $v_{0}$ on a bare piezoelectric substrate and the effective velocity $v_{e}$. The effective velocity is a result of perturbation by, for instance, the metal strips in the resonator and relates to the free velocity as $v_e=v_0+\Delta v_e+\Delta v_m$. This assumes that the electrical loading $\Delta v_e$ and the mechanical loading $\Delta v_m$ are independent. 

Although comprehensive information is available on suitable materials for SAW devices at room temperature, new materials can become viable at low temperature. One example of this is ZnO, which although commonly used as a thin film transducer on nonpiezoelectric substrates such as sapphire, diamond or SiO$_2$/Si \cite{morgan2007, Weber}, is not viable as a bulk crystal substrate at room temperature due to a substantial electrical conductivity, which damps the SAWs. This problem disappears at low temperature \cite{Magnusson2014}, making a very low loss SAW substrate, as discussed in \secref{sec:Resonators}.

\subsection{Fabrication of SAW devices}
\label{sec:Fabrication}

%\mt{Peter: I think only a few sentences are needed here, focusing on what is special about SAW devices. Most of this is just standard cleanroom methodology.}

Regardless of the desired substrate material, SAW devices can be fabricated using standard lithographic techniques. The metal forming the IDT is typically aluminum since it is both very light (which prevents mass loading effects) and an excellent superconductor at low temperatures. If top contact is to be made to the aluminum, a thin layer of palladium can be deposited on top to prevent formation of aluminum oxide. 

A condition for operating in the quantum regime is $\hbar \omega_{IDT} \gg k_B T$, where $T$ is typically $\unit[10-50]{mK}$ for a dilution refrigerator. This implies that $f_{IDT}$ must be on the order of GHz, and considering that SAW velocities are typically around $\unit[3000]{m/s}$ the required wavelength is below one micron. An IDT emits SAWs efficiently when the finger distance $p$ is half a wavelength and therefore lithography on the submicron scale is needed. These dimensions are difficult to reach with photolithography, so electron-beam lithography is typically used. In principle, etching could be used to fabricate the IDT but care would need to be taken to ensure that the surface where the SAW propagates is not damaged by the etching process. A lift-off process avoids the danger of possible surface damage, but contamination from remnants of resist is a concern.

%To do lift-off, the substrate is spin-coated with a thin layer of resist sensitive to electron radiation and patterned by electron-beam lithography. When lithography defines the pattern, it triggers a reaction in the resist. This reaction makes it solvable in a developer. The development opens up holes in the resist revealing the bare substrate according to the desired pattern. 
%After development, aluminum is deposited by evaporation in a high vacuum chamber. The aluminum source is heated by a current through a conductor (thermal evaporation) or by bombarding the source with electrons (electron beam evaporation). When the source has reached a certain temperature the metal evaporates and covers the entire substrate. 
%
%The evaporated metal film covers both the resist areas and the areas of exposed substrate where the resist was removed during the development. When the substrate is immersed into a remover bath, the resist is dissolved and removes the metal on top of the resist.  On the areas where the resist is absent, the metal sticks to the substrate. The resist is developed with an undercut to keep the metal from sticking to the side walls, allowing very sharp features to be achieved.  The metal remaining on the surface is the IDT.

\section{Theory}
\label{sec:Theory}

\subsection{Classical IDT model}
\label{sec:ClassicalTheory}

\begin{figure}[t!]
\centering
\includegraphics[width=0.6\linewidth]{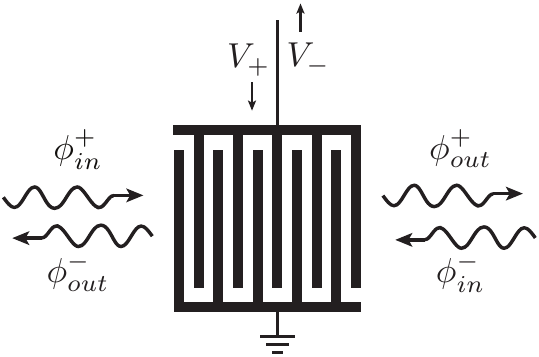}
\caption{An illustration of the three-port model for an IDT, featuring one electrical port and two acoustic ones. The IDT shown here has a single-finger structure. \label{fig:Sparams}}
\end{figure}

An IDT can be considered a three-port electrical device with two acoustic channels (rightward and leftward going waves are represented by Ó+Ó and Ó-Ó superscripts, respectivelyÓ), as shown in \figref{fig:Sparams}. Thus it can be described by a scattering matrix equation:
\begin{equation}
\begin{pmatrix}
\phi_{out}^- \\
\phi_{out}^+ \\
V^-
\end{pmatrix}
= 
\begin{pmatrix}  S_{11} &   S_{12} &   S_{13}\\
                           S_{21} &  S_{22} &  S_{23} \\
                           S_{31} &  S_{32} &  S_{33} 
\end{pmatrix}
\begin{pmatrix}
 \phi_{in}^+ \\
 \phi_{in}^- \\
 V^+
 \end{pmatrix}.
 \label{eq:t}
 \end{equation}
Here, $\phi_{in/out}^{\pm}$ represents the complex amplitude of an incoming (outgoing) SAWs on the left (right) side of the IDT and $V^{\pm}$ represents the complex amplitude of the incoming (outgoing) voltage wave on the IDT electrodes. Assuming reciprocity (a SAW travelling through the device left to right is given by time reversing a SAW travelling in the opposite direction) gives $S_{12}=S_{21}$, $S_{13}=S_{31}$, and $S_{23}=S_{32}$ (receiving a SAW is the time reversal of emitting a SAW). Assuming symmetry (the IDT looks the same to a SAW regardless of whether the SAW travels to the right or to the left) we have $S_{13}=S_{23}$, $S_{11}=S_{22}$, and $S_{31}=S_{32}$. In some cases, one might also be able to assume power conservation, in which case S would be unitary as well. 

Using just symmetry and reciprocity leaves four independent terms in S: $S_{11}$, $S_{12}$, $S_{13}$, and $S_{33}$. $S_{33}$ is the electrical reflection coefficient for a drive tone arriving at the IDT from the electrical transmission line
% $S_{3 3}=(Y_S-Y)/(Y_S+Y)$, where $Y$ is the admittance of the IDT and $Y_S$ is the admittance of the voltage source driving the IDT. 
$S_{11}$ is the reflection of the SAW off the IDT; it is a combination of the pure mechanical reflection and outgoing SAW regenerated by the voltage induced by the incoming SAW. $S_{12}$ is the transmission of SAW through an IDT, which again has both a mechanical component and a voltage regeneration component. $S_{13}$ represents the electro-acoustic transduction and is thus proportional to the transmitter response function $\mu$, which will be discussed below. 
%The form of $\mu$ is derived for a lossless IDT without mechanical reflections in \eqref{eq:Mu}. While the components of S are generally complicated to solve for, in \eqref{eq:Phi_em} we show that in the case of infinite load impedance and no mechanical reflections $S_{1 1}=-1$ and $S_{1 2}=0$.  

The simple single-finger form of the IDT, shown in \figref{fig:Sparams}, suffers from internal mechanical reflections which complicates the response and makes the scattering matrix only possible to estimate numerically, \textit{e.g.}, using the techniques known as the  Reflective Array Method (RAM) and Coupling Of Modes (COM) \cite{morgan2007}. Fortunately, one can eliminate these internal reflections using the double-finger structure, where each finger in the single-finger structure is replaced by two. The spacing between these two fingers can then be chosen such that reflection from the first finger interferes destructively with the reflection from the second finger. We will thus proceed with ideal single-finger and double-finger structures, assuming no mechanical reflection and no loss. This is approximately true for a superconducting double-finger structure but not always a good approximation for the single-finger structure.% (except possibly for IDTs with very few fingers). %These assumptions imply $P_{1 1}=P_{2 2}=0$ and $P_{1 2}=P_{2 1}=\exp(-jkl)$, a simple phase factor representing transversal of the SAW through the IDT without loss.

Following Datta \cite{datta1986}, the IDT is assumed to have a transmitter function $\mu(f)$ such that the emitted surface potential wave with amplitude $\phi_{em}$ is given by $\phi_{em}=\mu  V_t$ when a voltage of amplitude $V_t=V^+-V^-$ with frequency $f$ is applied. Likewise, the IDT has a receiver response function $g(f)$ such that an incoming SAW induces a current $I=g \phi_{in}$.
We define a characteristic impedance $Z_0$ such that a SAW of voltage amplitude $\phi$ carries power $P_{SAW}$ where $Z_0=1/Y_0=\left| \phi^2 \right| /2P_{SAW}$. It can be shown using reciprocity that $g=2 \mu Y_0$. %It is useful to define an equivalent transmission line model for SAW using the surface wave velocity $v_f$. 

The inductance and capacitance per unit length of the equivalent transmission line model for SAWs are given by the usual expressions $C=1/(Z_0 v_0)$ and $L=Z_0/v_0$, where $v_0$ is the free surface wave velocity. Since $v_0=1/\sqrt{LC}$, small changes in velocity are related to small changes in capacitance: $\left |\Delta v_0 /v_0 \right | = \Delta C/(2C)$. In a conducting metal film on top of the SAW substrate, the surface potential $\phi$ induces a surface charge density $q_s$ resulting in $\Delta C=-q_s W/\phi$ for a film of width $W$ (the effective length of the IDT finger) and a corresponding $\Delta v_0$. The resulting connection between applied voltage and SAW is called the piezoelectric coupling constant $K^2$, defined such that $2\Delta v_0/v_0=K^2$.
%where $\Delta v_0$ is the difference in speed for a SAW propagating on a free surface versus underneath a thin conducting metal film (thin enough that it doesn't contribute any mass loading). 
$K^2$ is a material property and has been calculated and measured for a variety of SAW substrates, see Table\,\ref{tab:MaterialProperties}. By considering the discontinuity caused by the surface charge density and using Maxwell's laws, it can be shown that % $K^2 y_0 = 2 \pi C_s v_0$
$Y_0=\omega_{IDT} W C_S /K^2$.
%\mc{Maria, you suggested that we should use the notation with $\epsilon$ instead, can you do this consistently, and modify Table 1 accordingly, make sure it agrees with the Oxford notation in Eq.\,(\ref{eq:6})} 

\begin{figure}[t!]
\centering
\includegraphics[width=0.85\linewidth]{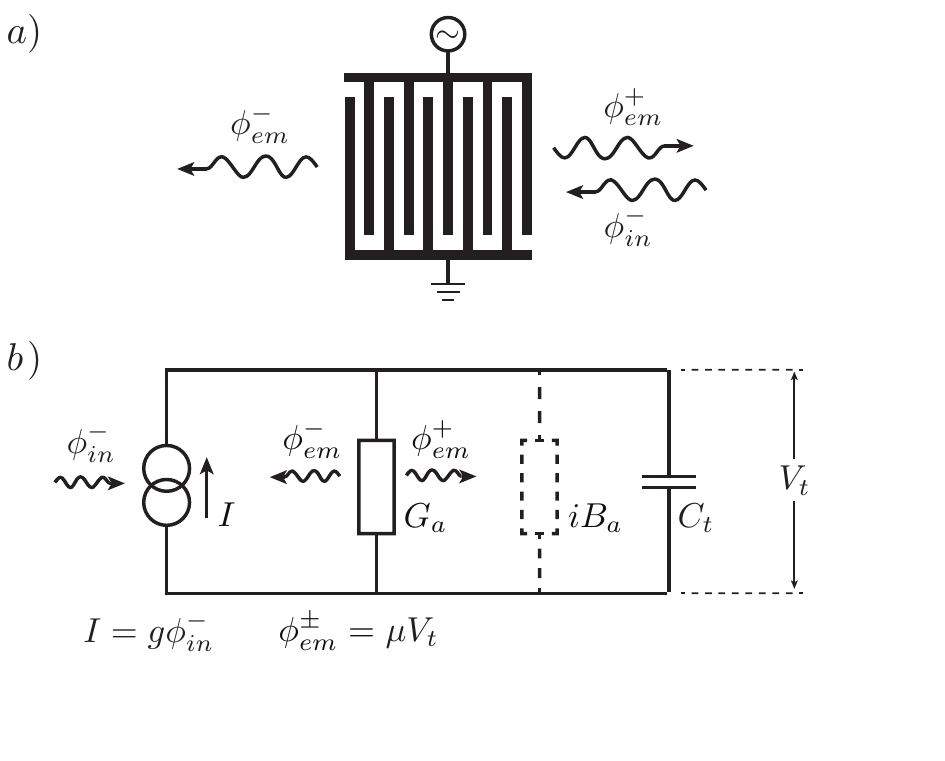}
\caption{Classical model for the IDT. (a) Layout of the IDT, with incoming and outgoing SAW components shown. (b) Equivalent circuit, see text.
\label{fig:IDT}}
\end{figure}

Without piezoelectric effects, the IDT is just a geometric capacitor $C_t$ with admittance $i\omega C_t$. The presence of piezoelectricity adds a complex admittance element $Y_a(\omega)=G_a(\omega)+i B_a(\omega)$ \cite{datta1986, morgan2007, campbell1998}, capturing the electro-acoustic properties. Including a current source representing the transduction form SAW to electricity, we have the circuit model for a receiver IDT, shown in \figref{fig:IDT}.
%By applying this model to an IDT with load $Z_L=1/Y_L$ (which in the case of a qubit corresponds to the shunting SQUID), we get the equivalent circuit of a receiving IDT shown in \figref{fig:IDT}. 
We can also consider a voltage source with the characteristic impedance $Z_C=1/Y_C$ and get the equivalent circuit for a transmitting IDT. When a voltage $V_t$ is applied to an emitting IDT, dissipation of electrical power in $G_a$ represents conversion into SAW power $P=\frac{1}{2} |V_t|^2 G_a$. This power is divided equally between the two directions of propagation. %$P=\dfrac{1}{2}\phi I^*=\dfrac{|\phi|^2}{2Z_0}$. 
By equating the electrical power to the SAW power, one derives $ G_a=2|\mu|^2 Y_0$. Because $\mu$ and $g$ 
%\mc{here we use $g$ and not $g_m$ since that is used in the quantum theory section}
are always purely imaginary, this can also be written as $G_a=-\mu g$. Matching $1/G_a$ to the $\unit[50]{\Omega}$ impedance line is important for the transmitting/receiving IDT to maximize signal and minimize electrical reflection since $S_{3 3}=(Y_C-Y)/(Y_C+Y)$ where $Y=G_a(\omega)+i B_a(\omega)+i \omega C_t$.
%\mc{ we use $i$ (not $j$ as $\sqrt{-1}$)}

The imaginary component of the acoustic admittance $B_a$ is the Hilbert transform of $G_a$ due to causality. Thus specifying $\mu$ and the capacitance $C_t$ yields the entire electro-acoustic behavior of the IDT. 
It can be shown that $\mu$ depends on the Fourier transform of the surface charge density, though this is either a complicated algebraic formula for regular (evenly spaced) transducers or must be evaluated numerically for nonregular transducers. Likewise the capacitance relates to the surface charge induced by an applied voltage, and is again a complicated algebraic formula for regular transducers, for more complicated structures it needs to be evaluated numerically. In the particular case of single-finger and double-finger electrodes with metallization ratio of $\unit[50]{\%}$, the results are
\bea
\mu_0^{sf} &=& 1.6 i \Delta v_0/v_0 \approx 0.8 i K^2, \quad\\
\mu_0^{df} &=& 1.2 i \Delta v_0/v_0 \approx (0.8/\sqrt{2}) i K^2, \quad \nn \\
C_t^{sf} &=& N_p C_S W, \quad\\
C_t^{df} &=& \sqrt{2} N_p C_S W.\nn
\eea
Here, $\mu_0$ is the response of one finger pair when all other fingers are grounded, and $C_t$ is the capacitance of an IDT with $N_p$ finger periods. Using superposition, this allows separating out the response of a single finger (called an element factor) from the effect of superimposing several fingers (the array factor) for regular transducers. 

The array factor is the sum of the phase factors from all the different fingers
\be
A(\omega)^{sf} = \sum_{n=0}^{N_p-1} \exp\left( -i 2 \pi n \frac{\omega}{\omega_{IDT}}\right),
\ee
%
%$$A(\omega)^{df} = \sum_{n=0}^{N_p-1} \exp(-2i k N_p)$$
which can be shown by a geometric series and small-angle approximation argument to yield 
\bea
A(\omega)^{sf} &=& N_p \frac{\sin(X)}{X}, \quad A(\omega)^{df} = \sqrt{2} N_p \frac{\sin(X)}{X}\\
X &=& N_p \pi \frac{\omega-\omega_{IDT}}{\omega_{IDT}}
\eea
In the approximation that the element factor for the double-finger case is smaller by roughly a factor of $\sqrt{2}$ while the array factor is greater by a factor of $\sqrt{2}$, the two cancel and we get the same form for the total response function of double-finger and single-finger IDTs:
\be
\mu= 0.8i K^2 N_p \frac{\sin(X)}{X}.
\label{eq:Mu}
\ee
% Assuming no loss, $G_a=$
Thus, we get that
\bea
G_a\left(\omega \right) &=& G_{a0} \left[\frac{\sin(X)}{X}\right]^2,\\
B_a\left(\omega \right) &=& G_{a0} \left[\frac{\sin(2X)-2X}{2X^2}\right],
\eea
where $G_{a0} \approx 1.3 K^2N_p^2\omega_{IDT} W C_S$.
%The more accurate results not making an approximation about the cancellation of the factor of $\sqrt(2)$ are:
%$$ G_a(\omega_{IDT})^{sf} \approx =1.44 K^2N_p^2\omega_{IDT} W C_S$$
%$$ G_a(\omega_{IDT})^{df} \approx =1.56 K^2N_p^2\omega_{IDT} W C_S$$
On acoustic resonance, $\omega=\omega_{IDT}$, $G_a$ is at its maximum and the imaginary element $B_a$ is zero.
% Off resonance $G_a=G_a(\omega_{IDT}) [\sin(X)/X]^2 $	with $X=N_{tr} \pi (\omega -\omega_{IDT})/\omega_{IDT}$ where $B_a$ is the Hilbert transform of $G_a$. 
The sinc function dependence of $\mu$ on frequency implies a bandwidth of approximately $0.9 f_{IDT}/N_p$.

%To semi-classically estimate the coupling of the qubit to SAW, we limit the discussion to the case of acoustic resonance so $\mu= 0.8 i \,K^2 N_{tr}$.

\subsection{Semiclassical theory for SAW-qubit interaction}
\label{sec:SemiClassicalTheory}

\begin{figure}[t!]
\centering
\includegraphics[width=0.9\linewidth]{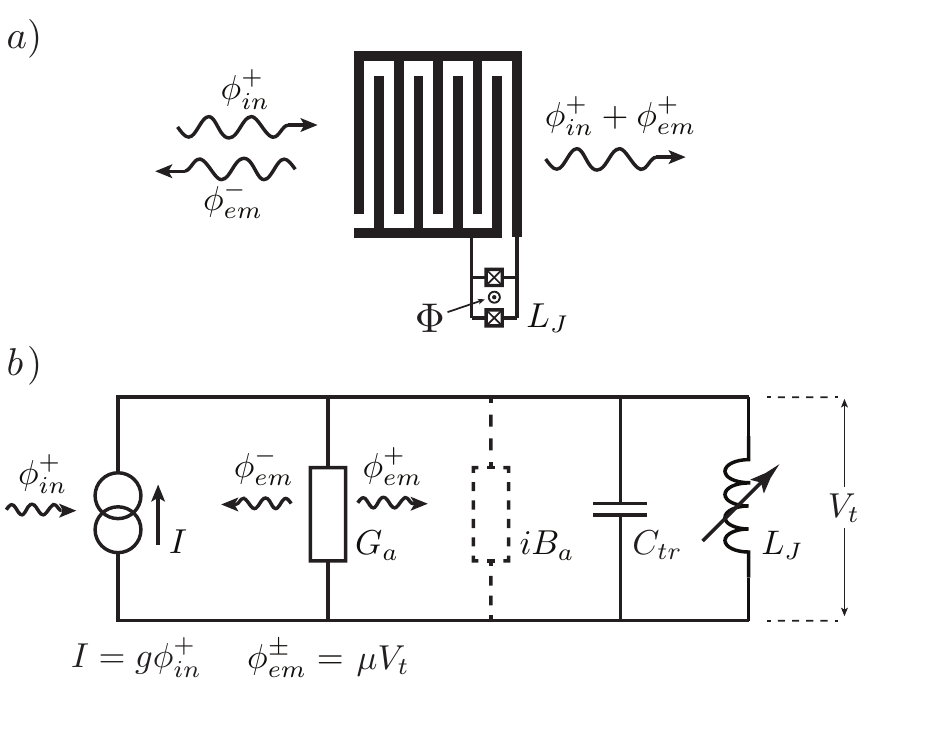}
\caption{Semiclassical model for the acoustically coupled qubit. (a) Layout of the qubit, with incoming and outgoing SAW components shown. The capacitively coupled gate is not included in this model. (b) Equivalent circuit, see text.\label{fig:classical_coupling}}
\end{figure}

In the experiment coupling a transmon qubit to SAWs \cite{Gustafsson2014}, we take advantage of the similarities between the interdigitated structure of a classical IDT and that of the transmon. The transmon consists of a SQUID connecting two superconducting islands that form a large geometric capacitance $C_{tr}$ in an interdigitated pattern. The SQUID acts as a nonlinear inductance $L_J$ and forms an electrical resonance circuit together with $C_{tr}$. The nonlinearity of $L_J$ gives the transmon an anharmonic energy spectrum. In this section, we consider a semiclassical model, valid when the incoming SAW power is low enough that the qubit is never excited beyond the $\ket{1}$ state. In that case, the SQUID can be approximated as a linear inductance. We can then use the circuit model shown in \figref{fig:classical_coupling}.

%In the experiment coupling a transmon qubit to SAWs \cite{Gustafsson2014}, we take advantage of the similarities between a classic IDT and the interdigitated structure of the two superconducting islands of the transmon. In addition to the usual IDT conductance $G_a$ and  capacitance $C_{tr}$, the transmon qubit includes a SQUID, shunting the two electrodes as shown in \figref{fig:classical_coupling}. The SQUID acts as a non-linear inductance $L_J$ and forms an electrical resonance circuit together with $C_{tr}$. For the semiclassical model of this section to be valid, the incoming SAW power must be low enough that the qubit is never excited beyond the $\ket{1}$ state. In that case, the SQUID can be approximated as a linear inductance.

Lithography fixes the acoustic resonance frequency $\omega_{IDT}$, but the electrical resonance frequency of the transmon can be tuned by adjusting $L_J$ with a magnetic field. The two resonances coincide when
\be
L_J=\dfrac{1}{\omega_{IDT}^2 C_{tr}}.
\ee
At this point, the impedance of the $LC$ circuit approaches infinity, and we are left with only the acoustic element $G_a$. As a result, the current generated by an incoming SAW beam with amplitude $\phi_{in}^+$ in the rightward direction produces a voltage over $G_a$,
\be
V_t=I/G_a=g \phi_{in}^+/G_a,
\ee
which in turn gives rise to re-radiation of SAW in the rightward and leftward directions with amplitudes
\be
\phi_{em}^{\pm} = \mu V_t = (\mu g/ {G_a}) \phi_{in}^+ = -\phi _{in}^+.
\label{eq:Phi_em}
\ee
Hence, the net transmission of SAW in the rightward direction is $\phi_{out}^+ = \phi_{in}^+ + \phi_{em}^+=0$, and the
emission in the leftward direction is	$\phi_{out}^-=-\phi_{in}^+$. This explains the acoustic reflection $S_{1 1}$ we observe experimentally (see \secref{sec:SAWqbInteraction} for details) in the limit of low SAW power.

The semiclassical model can also be used to estimate the relaxation rate (coupling) $\Gamma_{ac}$ of the qubit to phonons. The damping rate of the $RLC$ circuit tells us how fast electrical energy stored in the $LC$ resonator converts into SAWs by dissipating in $G_a$, giving
\be
\Gamma_{ac} = \omega_{IDT} \dfrac{G_a}{2} \sqrt{\dfrac{L_J}{C_{tr}}}=\dfrac{G_a}{2C_{tr}}.
\ee
%
%Using Eqs. (1), (3), (5), (6) and (7), we get
Thus, using our expressions for $G_{a0}$ and $C_{tr}$ from \secref{sec:ClassicalTheory}, we get a simple expression for the coupling between the qubit and the SAWs:
\be
\Gamma_{ac}=\dfrac{1.3 \,K^2N_p\omega_{IDT}}{ 2\sqrt{2} } \approx 0.5 \,K^2N_p\omega_{IDT}.
\label{eq:classical_coupling}
\ee
%
%=\dfrac{\omega_{IDT} 0.8^2 K^2 N_p}{\sqrt{2}}=2 \pi \times 30 MHz$$

\subsection{Quantum theory for giant atoms}
\label{sec:QuantumTheory}

To go beyond the semiclassical model and understand the behaviour of the transmon coupled to SAWs in regimes where stronger drive strengths are used and more levels of the artificial atom come into play, a fully quantum description is needed. From a quantum optics perspective, one of the main reasons that the transmon coupled to SAWs is a very interesting system is that it forms a ``giant artificial atom''. Natural atoms used in traditional quantum optics typically have a radius $r\approx \unit[10^{-10}]{m}$ and interact with light at optical wavelengths $\lambda \approx \unit[10^{-7}-10^{-6}]{m}$ \cite{LeibfriedRMP2003, MillerJPB2005}. Sometimes the atoms are excited to high Rydberg states ($r\approx \unit[10^{-8}-10^{-7}]{m}$), but they then interact with microwave radiation ($\lambda \approx \unit[10^{-3}-10^{-1}]{m}$) \cite{HarocheRMP2013, WaltherRPP2006}. Microwaves also interact with superconducting qubits, but even these structures are typically measured in hundreds of micrometers (although some recent designs approach wavelength sizes \cite{KirchmairNature2013}). Consequently, theoretical investigations of atom-light interaction usually rely on the dipole approximation that the atom can be considered point-like when compared to the wavelength of the light. This is clearly not the case for the transmon coupled to SAWs, since here each IDT finger is a connection point and the separation between fingers is always on the order of wavelengths.

\begin{figure}[t!]
\centering
\includegraphics[width=\linewidth]{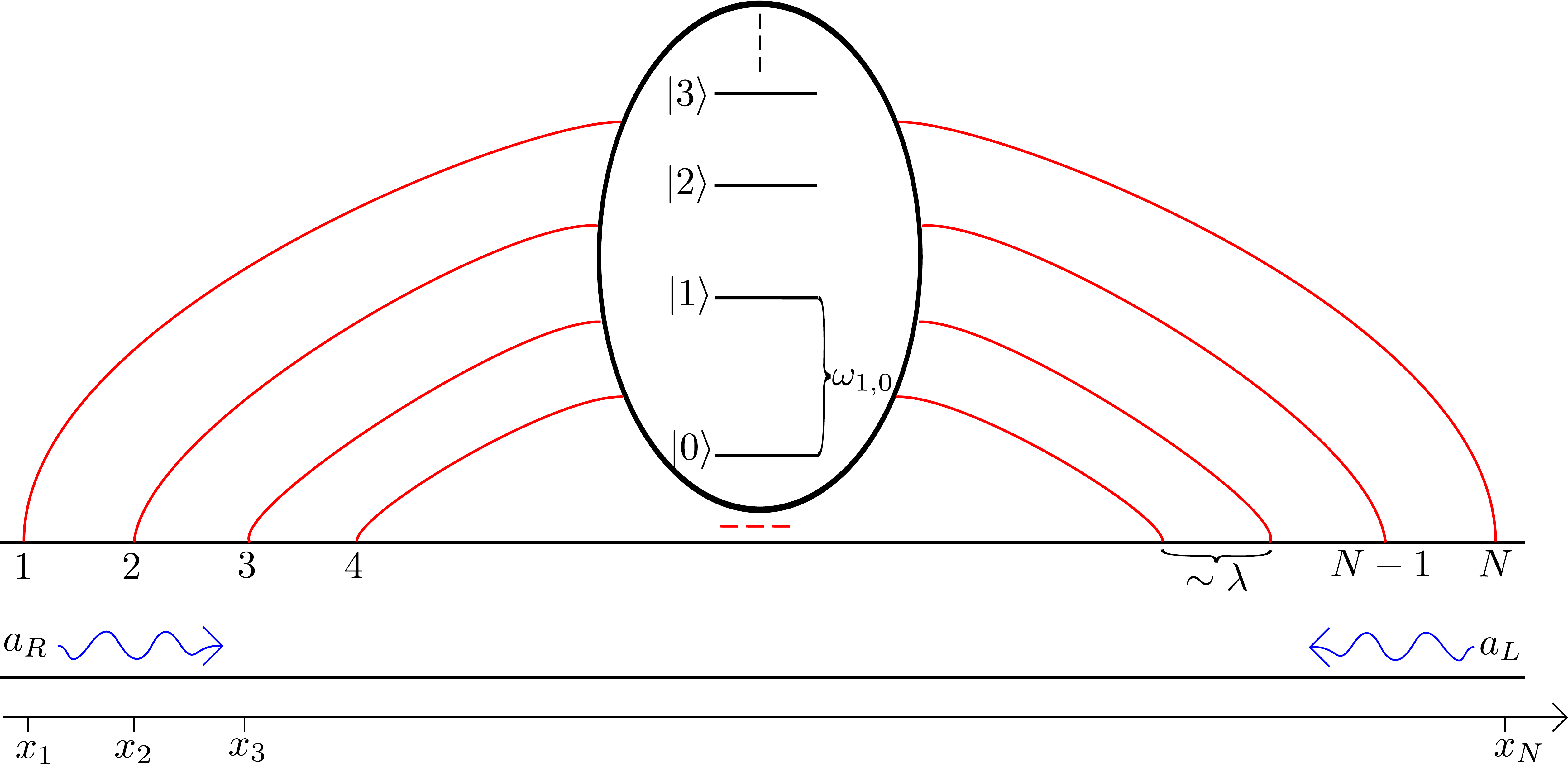}
\caption[A model for a giant artificial atom]{The quantum model for a transmon coupled to SAWs: a giant artificial multi-level atom, connected at $N$ points to left- and right-moving excitations in a 1D transmission line. \label{fig:GiantAtomH}}
\end{figure}

Conceptually, the SAW-transmon setup is equivalent to a model where an atom couples to a 1D continuum of bosonic modes at a number of discrete points, which can be spaced wavelengths apart. Such a setup should also be possible to realize with a variation of the transmon design, the ``xmon'' \cite{BarendsPRL2013}, coupled to a meandering superconducting transmission line for microwave photons. In Ref.~\cite{KockumPRA2014}, we investigated the physics of this model, a sketch of which is shown in \figref{fig:GiantAtomH}. Here, we summarize the main results from that paper.

The Hamiltonian for our model is
\bea
H = H_a + H_{tl} + H_{int},
\eea
where we define the multi-level-atom Hamiltonian
\be
H_a = \sum_m \hbar \omega_m \ketbra{m}{m},
\ee
the transmission line Hamiltonian
\be
H_{tl} = \sum_j \hbar \omega_j \left(a^\dag_{Rj} a_{Rj} + a^\dag_{Lj} a_{Lj} \right),
\ee
and the interaction Hamiltonian
\bea
H_{int} &=& \sum_{j,k,m} \hbar g_{jkm} \left(\sm^m + \sp^m \right) \times \\
&&\left(a_{Rj} e^{-i\frac{\omega_j x_k}{v_0}} + a_{Lj} e^{i\frac{\omega_j x_k}{v_0}} + \text{H.c.} \right),\nn
\eea
where $\sm^m = \ketbra{m}{m+1}$, $\sp^m = \ketbra{m+1}{m}$, and $\text{H.c.}$ denotes Hermitian conjugate. The atom has energy levels $m= 0,1,2, \dots$ with energies $\hbar \omega_m$. It is connected to right- and left-moving bosonic modes $Rj$ and $Lj$ of the transmission line with some coupling strength $g_{jkm}$ at $N$ points with coordinates $x_k$. We assume that the time it takes for a transmission line excitation to travel across all the atom connection points is negligible compared to the timescale of atom relaxation. Thus, only the phase shifts between connection points need to be included in the calculations, not the time delays. In addition, we assume that the coupling strengths $g_{jkm}$ are small compared to the relevant $\omega_m$ and $\omega_j$ and that they can be factorized as $g_{jkm} = g_j g_k g_m$, which is the case for the transmon \cite{Koch}. In general, the mode coupling strength $g_j$ can be considered constant over a wide frequency range. The factors $g_k$ are dimensionless and only describe the relative coupling strengths of the different connection points. Finally, for the transmon \cite{Koch} and other atoms with small anharmonicity, we have $g_m = \sqrt{m+1}$.

Using standard techniques, including the Born, Markov, and rotating-wave approximations \cite{CarmichaelBook1,GardinerZoller}, we derive the master equation for the effective density matrix of the atom, $\rho$. The result, assuming negligible temperature, is
\bea
\dot{\rho}(t) &=& - i \comm{\sum_m \left(\omega_m + \Delta_m \right)\ketbra{m}{m}}{\rho(t)} + \\
&&\sum_m \Gamma_{m+1,m} \lind{\sm^m}\rho(t),  \nn
\label{eq:MasterEq}
\eea
where $\hbar\Delta_m$ are small shifts of the atom energy levels and we use the notation $\lind{O}\rho = O\rho O^\dag - \frac{1}{2} O^\dag O\rho - \frac{1}{2} \rho O^\dag O$ for the Lindblad superoperators \cite{Lindblad1976} that describe relaxation. The relaxation rates $\Gamma_{m+1,m}$ for the transitions $\ket{m+1}\rightarrow\ket{m}$ are given by
\be
\Gamma_{m+1,m} = 4\pi g_m^2 J(\omega_{m+1,m}) \abssq{A(\omega_{m+1,m})}, \label{eq:DefGammam}
\ee
where $A(\omega_j) = g_j \sum_k g_k e^{i\omega_j x_k/v_0}$ contains the array factor from the classical SAW theory above, $J(\omega)$ is the density of states for the bosonic modes, and $\omega_{r,s} = \omega_{r} - \omega_{s}$. $A(\omega)$ enters squared, just like $G_\text{a}\propto \abssq{\mu}$; this gives a frequency-dependent coupling set by interference between the coupling points. Thus, we can design our artificial atom such that it relaxes fast at certain transition frequencies, but remains protected from decay at others. 

The shift of the atom energy levels by $\hbar\Delta_m$ in \eqref{eq:MasterEq} is an example of a Lamb shift \cite{LambPR1947, BethePR1947}, which is a renormalization of the atom energy levels caused by the interaction with vacuum fluctuations of the 1D continuum. The shifts are given by
\bea
\Delta_m &=& 2\mathcal{P} \int_0^\infty \id \omega \frac{J(\omega)}{\omega}\abssq{A(\omega)} \times \\
&&\bigg( \frac{g_m^2 \omega_{m+1,m}}{\omega+\omega_{m+1,m}} - \frac{g_{m-1}^2 \omega_{m,m-1}}{\omega-\omega_{m,m-1}} \bigg), \nn\label{eq:DefDeltamRenorm}
\eea
where $\mathcal{P}$ denotes the principal value.

Again, this is essentially equivalent to the imaginary acoustic admittance $iB_\text{a}$, which shifts the $LC$ resonance frequency in the semiclassical calculation above if the atom is not on electrical resonance with the IDT structure.

In conclusion, the giant artificial atom differs from an ordinary, ``small'' atom in that both its relaxation rates and its Lamb shifts become frequency-dependent. The intuition for this frequency-dependence carries over from the classical SAW theory.

\section{SAW resonators for quantum devices}
\label{sec:Resonators}

\begin{figure}[t!] 
%\begin{minipage}[t]{0.5\columnwidth}%
%\includegraphics[width=\linewidth]{SAWres_v2.pdf}%
%\end{minipage}%
%\begin{minipage}[t]{0.5\columnwidth}%
%\includegraphics[width=\linewidth]{S11magphase_v2.pdf}%
%\end{minipage}
\centering
\includegraphics[width=\linewidth]{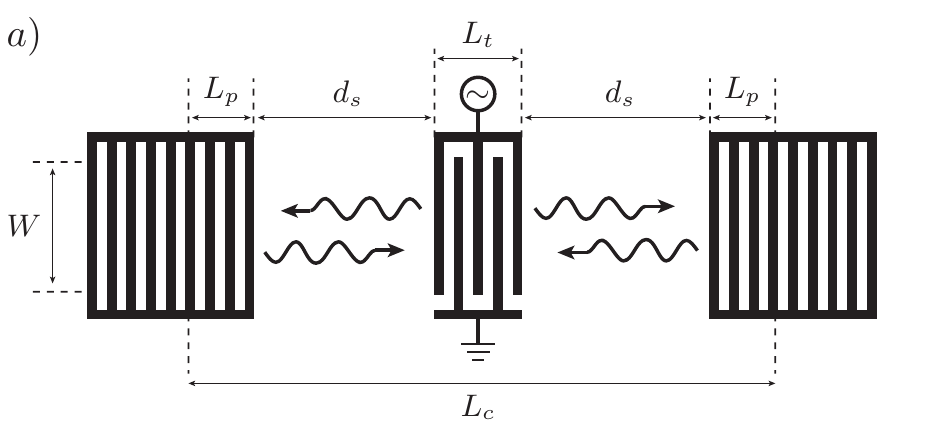} \\
\includegraphics[width=\linewidth]{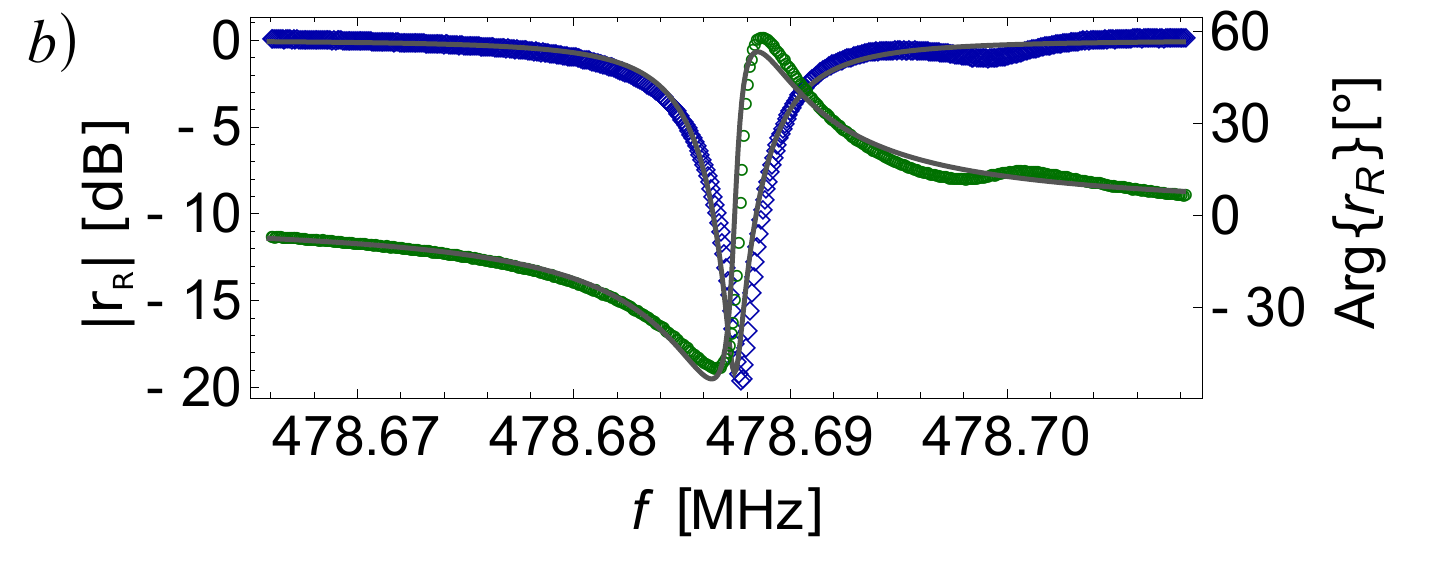}
\caption{(a) Schematic of a 1-port SAW resonator, with total effective cavity length $L_c$. (b) Measurement of the reflection coefficient $r_R$ of a 1-port SAW resonator on quartz at a wavelength of $p=\unit[3]{\mu m}$. Blue and green circles are the magnitude and phase of $r_{R}$ respectively, and the solid line is a fit to \eqref{eq:ReflectionCoeff}. \label{fig:ResonatorSchematic}}
\end{figure}

Seeing that a qubit can be coupled to SAWs, it is natural to consider the prospects for trapping SAW phonons in resonant structures, for example for implementation of a SAW phonon version of circuit QED. High quality SAW resonators have long been used in conventional SAW devices \cite{morgan2007}, for example to implement high quality oscillators, having been first introduced in the 1970s \cite{Ash1970}. A resonator is made by creating high-reflectivity mirrors for a SAW, which can be achieved using a shorted or open grating of many electrodes, with period $p=\lambda/2$ (see \figref{fig:ResonatorSchematic}). Such gratings operate in much the same way as a Bragg grating in optics, with constructive interference occurring between the multiple reflections from the electrodes when the device is on resonance. A signal can be coupled into and out of the device with an IDT, and the frequency response measured to obtain information about the resonator modes and their quality factors.

A schematic of a one-port SAW resonator and its frequency response are shown in \figref{fig:ResonatorSchematic}. The frequency response close to resonance can be modelled as a $RLC$ resonator to obtain the following formula for the reflection coefficient:
\begin{equation}
r_{R}\left(f\right)=\frac{\left(Q_e-Q_i\right)+2 i Q_i Q_e \Delta f/f_0}{\left(Q_e+Q_i\right)+2 i Q_i Q_e \Delta f/f_0}, \label{eq:ReflectionCoeff}
\end{equation}
where $f_0$ is the resonant frequency for the SAW cavity, $\Delta f=f-f_0$ and $Q_i$, $Q_e$ are the internal and external quality factors, respectively. Note that here the actual capacitance of the IDT is neglected.

\subsection{Resonator quality factors at low temperature}
The external quality factor $Q_e$ of a SAW resonator is determined by the IDT geometry and external circuit parameters, whereas the internal quality factor $Q_i$ has contributions from multiple sources, including diffraction, conversion to bulk phonons, and finite grating reflectivity and resistivity. In the following we describe preliminary measurements of SAW resonator quality factors on ST-cut quartz measured at low temperature, characterizing two of these contributions specifically - the $Q_i$ due to finite grating reflectivity, and $Q_e$ due to the IDT. In all cases, parameters are extracted from fits to \eqref{eq:ReflectionCoeff}, after appropriate calibration of imperfections from the measurement circuit.

\begin{figure}[t!]
\centering
\includegraphics[width=\linewidth]{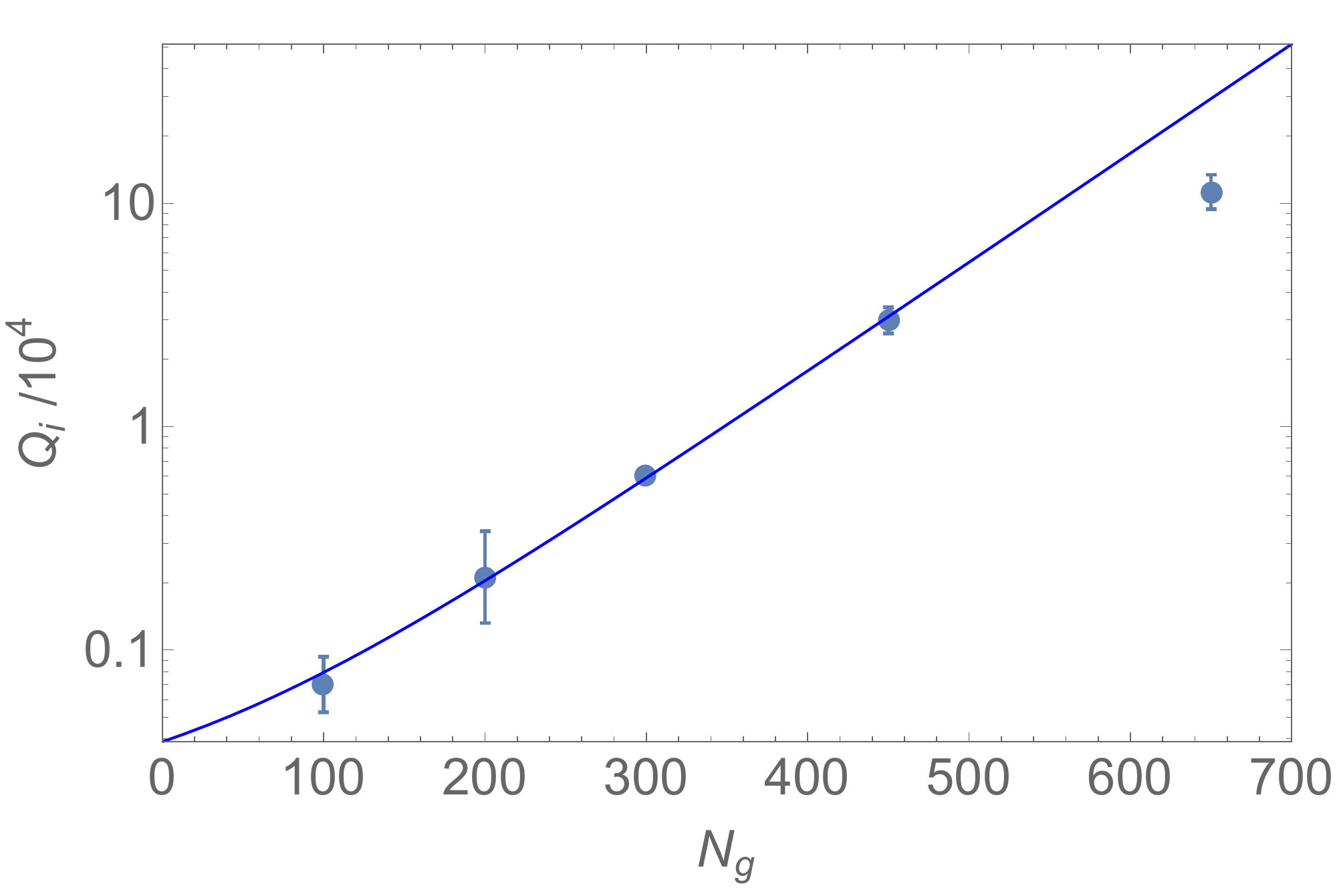}
\caption{Measurements of internal quality factor $Q_i$ of a set of SAW resonators on ST-cut quartz with $p=\unit[3]{\mu m}$, as a function of number of electrodes in the grating reflectors $N_g$, measured at $\unit[10]{mK}$. The blue line is a fit to \eqref{eq:Qd}.
}
\label{fig:NgQi}
\end{figure}

\figref{fig:NgQi} shows measurements of $Q_i$ for a set of devices with $p=\unit[3]{\mu m}$, the cavity width $W=750\,\mu$m, the cavity length $L_c\approx 460\,\mu$m ($f_0\approx \unit[0.522]{GHz}$). The number of grating electrodes $N_g$ is varied. The contribution to $Q_i$ due to the grating reflectivity, $Q_g$, can be derived by summing the multiple reflections arising from the grating electrodes, which in the limit of large $N_g$ is given by \cite{morgan2007}
\begin{equation}
Q_{g}(L_c,N_g) = \frac{\pi L_{c}}{\lambda_{0}\left(1-\tanh\left(\left|r_{s}\right|N_{g}\right)\right)}\label{eq:Qg},
\end{equation}
with $L_c$ the cavity length, $\lambda_0$ the cavity wavelength, and $r_s$ the reflectivity of a single grating electrode. The data fit extremely well to this equation,
% in the low $N_g$ limit (????)
indicating that the gratings indeed behave according to this simple model, and we can conclude that any other contributions to the dissipation are at the $Q_i>10^5$ level. Further experiments at larger $N_g$ will enable determination of contributions from other sources such as diffraction, which is expected to follow \cite{Bell1976}
\begin{equation}
Q_{d}(W) = \frac{5\pi}{\left|1+\gamma\right|}\left(\frac{W}{\lambda_0}\right)^{2}, \label{eq:Qd}
\end{equation}
where the diffraction parameter $\gamma=0.378$ for ST-X quartz \cite{Slobodnik1978}.

\begin{figure}[t!]
\centering
\includegraphics[width=\linewidth]{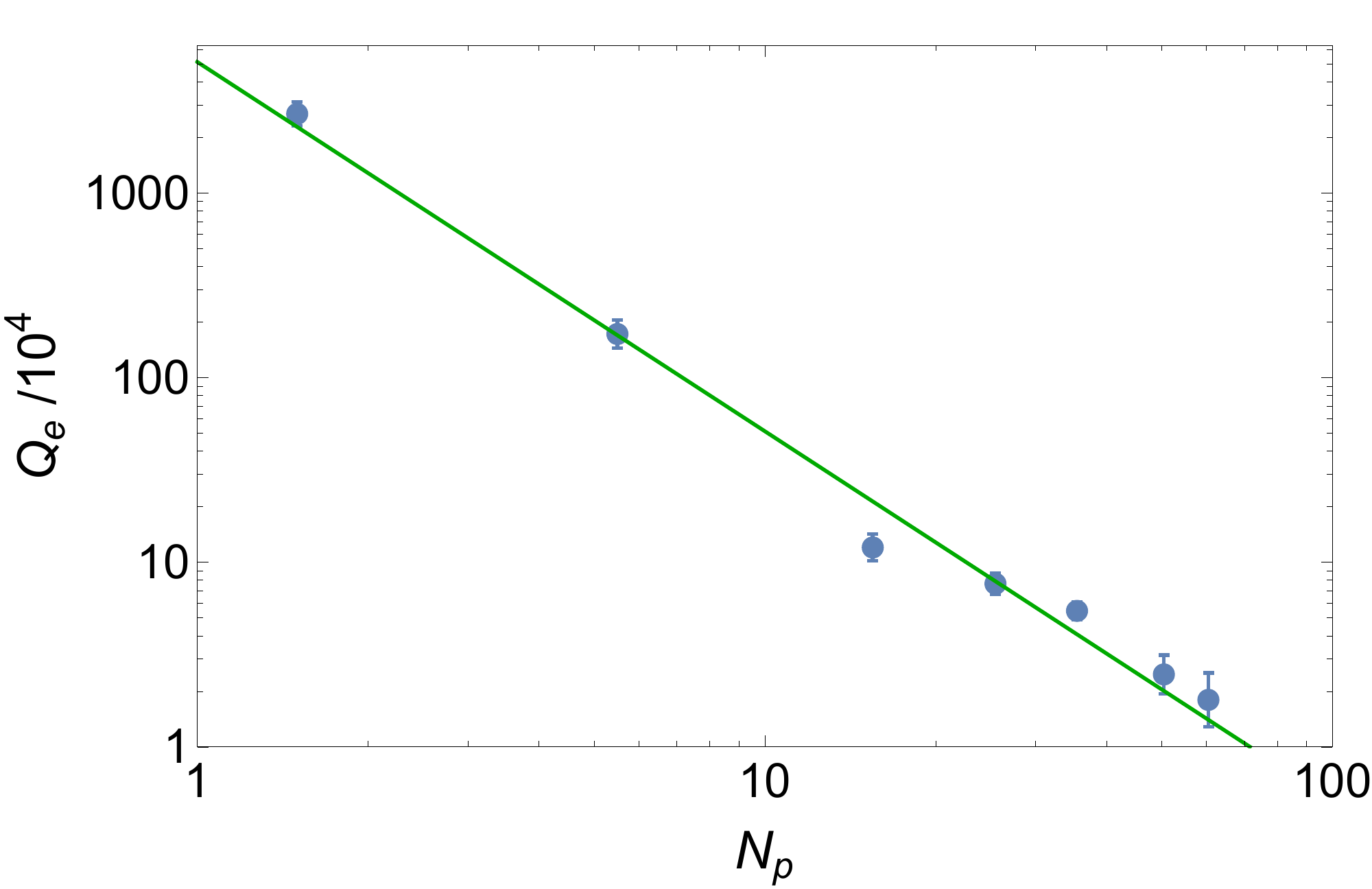}
\caption{Measurements of external quality factor $Q_e$ of SAW resonators on ST-cut quartz at $p=\unit[3]{\mu m}$, as a function of number of finger pairs in the IDT $N_p$, measured at $\unit[10]{mK}$. The green line is a fit to \eqref{eq:Qe}.}
\label{fig:NpQe}
\end{figure}

\figref{fig:NpQe} shows measurements of $Q_e$ for a set of devices with $p=\unit[3]{\mu m}$, $W=160\,\mu$m, and $L_c\approx 200\,\mu$m in which the number of electrodes in the IDT, $N_p$ is varied. A larger IDT naturally couples more strongly to the resonator, giving lower $Q_e$. A full expression for the expected dependence is given by \cite{morgan2007}
\begin{equation}
Q_{e}(L_c,N_p)=\frac{1}{5.74\,v_{0}Z_{C} C_S W K^2}\frac{L_{c}}{N_{p}^{2}},
\label{eq:Qe}
\end{equation}
where $Z_C$ is the characteristic impedance of the electrical port coupled to the IDT.
% and $\epsilon_{\infty}=\lim_{k\rightarrow\infty}\epsilon\left(k\right)$ is the effective permittivity. 
The data fit extremely well to this equation, which shows that a wide range of external quality factors, from $10^4$ to above $10^7$ can be engineered. Combining this information with the observed internal quality factors of up to $10^5$ shows that strongly under or over-coupled resonators can easily be fabricated.

%The variable $\beta$ is a phenomenological parameter and its value is $\beta=5.1$. 

\subsection{ZnO for high-Q SAW devices at low temperature}

\begin{figure}[t!]
\centering
\includegraphics[width=\linewidth]{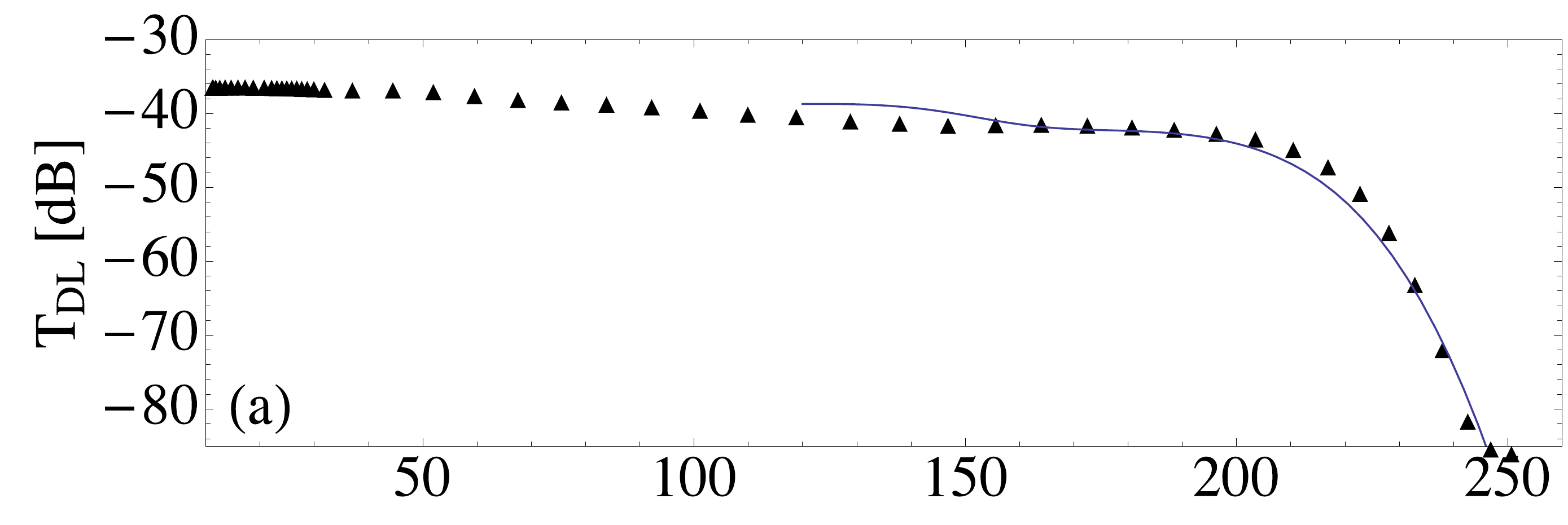} \\
\includegraphics[width=\linewidth]{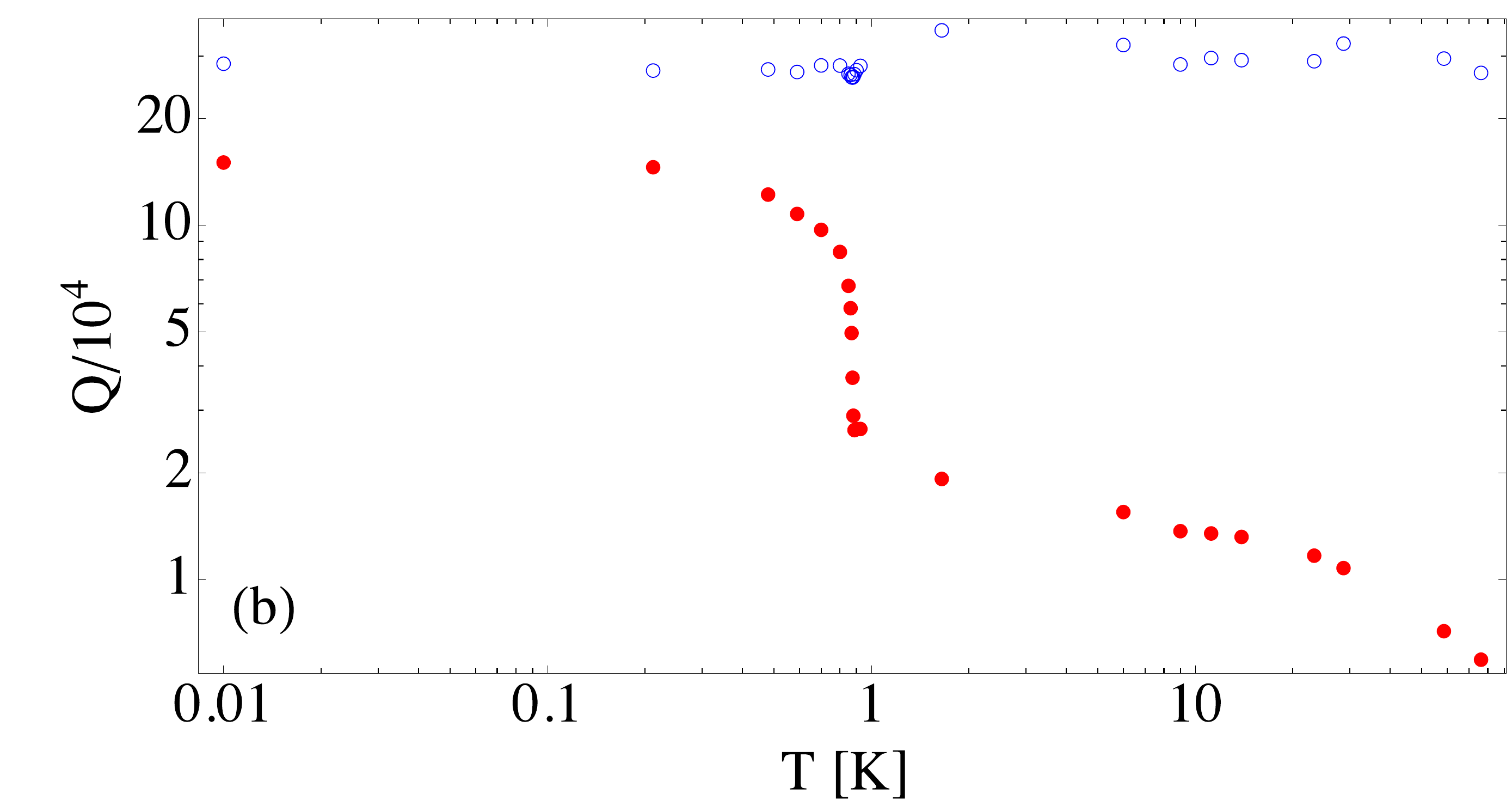}
\caption{\label{fig:ZnO} 
(a) Transmission amplitude $T_{DL}$ of the fundamental mode ($f=\unit[446]{MHz}$) of a ZnO delay line as a function of temperature $T$. The solid line is a fit of transmission attenuating linearly with the measured conductivity. (b) Temperature dependence of internal quality factor $Q_i$ (filled circles) and external quality factor $Q_e$ (empty circles) of a one-port SAW resonator on ZnO, at $f_0\simeq\unit[1.7]{GHz}$. Figures taken from \cite{Magnusson2014}. 
}
\end{figure}

Bulk ZnO has intrinsic carriers at elevated temperatures, and is thus only feasible as a material for SAW devices at low temperatures. In \cite{Magnusson2014}, results are reported on a delay line and a one-port resonator  fabricated on a $\unit[0.5]{mm}$ thick high quality ZnO substrate, measured in a dilution refrigerator down to $\unit[10]{mK}$. We summarize the important results in \figref{fig:ZnO}, in which we show the temperature dependence of the transmission of the delay line, and of the quality factor of the resonator. Remarkably, not only does the substrate become viable for these SAW devices at low temperature, but it also proves to have very low loss, with the resonator at $\unit[1.7]{GHz}$ reaching an internal quality factor of $Q_i \simeq 1.5 \times 10^5$ at $\unit[10]{mK}$.

The delay line device has a parallel IDT design with $p=\unit[2]{\mu m}$ and $p=\unit[3]{\mu m}$ transducers, and a mirrored output IDT at $\unit[2]{mm}$ separation.
%The fundamental modes are seen at $f_1=446.4$\,MHz and $f_2=669.5$\,MHz.
As \figref{fig:ZnO}(a) shows, the transmittance through the delay line (at the fundamental frequency $f_1$) is quenched around $\unit[200-240]{K}$ due to the onset of conductance. The line is a fit to the data of $f(T) = a \cdot e^{-b/\rho_e(T)}$, with parameters $a = \unit[-38.3]{dB}$ and $b = \unit[2.65\times10^8]{\Omega m}$ ($\rho_e(T)$ is the resistivity of the substrate). The agreement with the data demonstrates that the attenuation is indeed inversely proportional to the measured resistivity $\rho_e$, showing that this is the dominant source of loss in the high temperature range. The resonator has a single IDT with 21 fingers, and gratings of 1750 fingers on either side, with $p=\unit[0.8]{\mu m}$, with a resonance seen at $f_0\simeq\unit[1.677784]{GHz}$ at $\unit[10]{mK}$. As can be seen in \figref{fig:ZnO}(b), the internal quality factor drops by almost an order of magnitude in the range $0.01<T< \unit[1]{K}$, indicating a strong contribution from the superconductivity of the Ti/Al bilayer electrodes.

\section{SAW-qubit interaction in experiment}
\label{sec:SAWqbInteraction}

Having seen the experimental results for SAW resonators and the theory for IDTs and qubits, we now turn to an overview of the results presented in \cite{Gustafsson2014}, where the coupling between SAWs and a superconducting qubit was demonstrated. We highlight some of the technical considerations that apply to the design of hybrid quantum-acoustical devices.
%This section gives an overview of the results presented in \cite{Gustafsson2014}, where the coupling between SAWs and a superconducting qubit was demonstrated. We highlight some of the technical considerations that apply to the design of hybrid quantum-acoustical devices.

\begin{figure}[t!]
\begin{center}
\includegraphics[width=\linewidth]{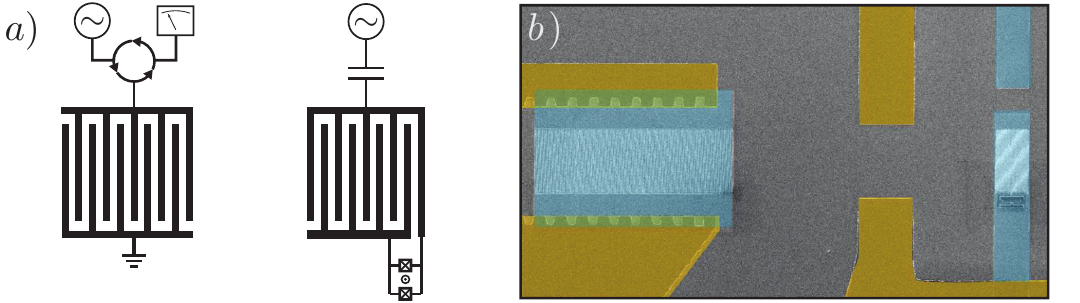}
\caption{(a) Simplified layout for the SAW-qubit sample. A surface acoustic wave is generated by the IDT and launched towards the transmon qubit, with its capacitance shaped into an IDT. This setup allows both to test SAW reflection on the qubit, and to listen to phonons emitted by the relaxing qubit. (b) False color picture of the sample. The bluish parts are the IDT (to the left) and the qubit with its gate (to the right). The yellow parts are the coplanar waveguide and surrounding ground plane}
\label{fig:Qubit}
\end{center}
\end{figure}

The sample used in \cite{Gustafsson2014} (see \figref{fig:Qubit}) consisted of a polished GaAs substrate. A single IDT on one end of the chip is used to convert electrical microwaves into SAWs and vice versa. The IDT is aligned with the $[011]$ direction of the crystal. $\unit[100]{\mu m}$ away from the IDT, the transmon qubit is deposited, with the shunt capacitance fashioned into a finger structure as described above, aligned with the IDT. When the IDT is excited electrically, it emits a coherent SAW beam in the direction of the qubit, and the phonons that are reflected or emitted by the qubit can be detected by the IDT. One of the qubit electrodes is grounded, and the other one couples to an electrical gate through a small capacitance.

Although both the IDT and the qubit have interdigitated structures, they are subject to different constraints, and this presents a challenge in the choice of materials and layout. To achieve optimal electro-acoustic conversion in the IDT, its impedance should be matched as well as possible to that of the electrical transmission line it is connected to, which is typically $\unit[50]{\Omega}$. This is generally easiest to achieve on a strongly piezoelectric substrate material such as LiNbO$_3$. In that case only a few finger periods are needed to bring the real part of the acoustic impedance, $G_0$, close to $\unit[50]{\Omega}$. This gives the IDT a large acoustic bandwidth and reduces the influence of internal mechanical reflections, as well as of the shunt capacitance $C_{IDT}$.

The acoustic impedance of the qubit, on the other hand, does not need to be matched to any electrical transmission line, since its coupling to the electrical gate is designed to be weak. The strength of the piezoelectric coupling constant and the number of finger periods determines the acoustic coupling rate $\Gamma_{ac}$, which represents the rate at which the qubit can absorb and emit phonons (see \eqref{eq:classical_coupling}). It is desirable that this rate exceeds any coupling to electromagnetic modes, and a high rate of phonon processing also relaxes the requirements on signal fidelity through the IDT and the amplifier chain.

An essential feature of a qubit, however, is that the transitions between its different energy levels can be separately addressed. The separation between transition energies, $a = \omega_{21}-\omega_{10}$ is known as the \emph{anharmonicity} of the qubit. If $\Gamma_{ac}$ approaches $\abs{a}$, a signal used to excite the qubit from the ground state $| 0 \rangle$  to the first excited state $| 1 \rangle$ is also capable of exciting the $| 2 \rangle$ state. This means that the qubit ceases to work as a two-level system. The transmon design is a good candidate for coupling to SAWs since its large shunt capacitance can be shaped into an IDT-like structure. However, this design also comes with inherently low anharmonicity \cite{Koch}, which puts an upper bound on $\Gamma_{ac}$. A straightforward way to reduce $\Gamma_{ac}$ is to lower the number of finger periods, $N_{tr}$, but the finger structure of the qubit needs to have at least a few finger pairs in order to couple preferentially to the desired Rayleigh modes. On a strongly piezoelectric material, it is not necessarily possible to achieve a good balance between coupling strength and anharmonicity.

In the sample discussed here, the trade-off between IDT and qubit performance was managed by using GaAs, which is only weakly piezoelectric ($K^2 \approx 7 \times 10^{-4}$). With a moderate number of finger periods in the qubit, $N_{tr} = 20$, spaced for operation around $\omega_{10}/2\pi = \unit[4.8]{GHz}$, this gives a coupling of $\Gamma_{ac}/2\pi = \unit[30]{MHz}$ according to \eqref{eq:classical_coupling}. The width of the SAW beam (length of the fingers) is $W = \unit[25]{\mu m}$ and the fingers are pairwise alternating, in a design that minimizes internal mechanical reflections \cite{Bristol1972,datta1986,morgan2007} (\emph{c.f.} \secref{sec:ClassicalTheory}). With this design, the acoustic bandwidth of the qubit substantially exceeds $\Gamma_{ac}$ and $C_{tr}$ is sufficiently high for the qubit to operate well in the transmon regime.

To match the IDT to $\unit[50]{\Omega}$ with the same kind of ideal (nonreflective) finger structure would, however, not be possible on this substrate. Several thousand fingers would be required to reach optimal impedance matching, which makes fabrication infeasible and introduces problems due to the high shunt capacitance. To achieve the required impedance matching, the IDT instead consists of a single-finger structure where internal mechanical reflections are prominent. These reflections confine SAWs within the finger structure, achieving a stronger coupling (lower impedance) with a manageable number of fingers. The optimal number of fingers in this configuration was found experimentally to be $N_p = 125$. This value depends on the mechanical reflection coefficient of each IDT finger, which in turn depends on the material and thickness of the metallization. The resonant operation of the IDT, along with the relatively high value of $N_p$ compared with $N_{tr}$, makes the bandwidth of the IDT very slim, $\sim \unit[1]{MHz}$. This is the band in which phonons can be launched toward the qubit and phonons emanating from the qubit detected.

The capacitive gate suffers no such bandwidth limitation, and can be used to address transitions in the qubit also outside the acoustic frequency band of the IDT. The coupling to the gate is engineered to be sufficiently weak that the excited qubit preferentially relaxes by emitting SAW phonons.

In the article by Gustafsson \textit{et al.}, three different experiments were presented (see \figref{fig:ThreeExp}).
The experimental data demonstrate the following key features:

\begin{figure}[t!]
\begin{center}
\includegraphics[width=\linewidth]{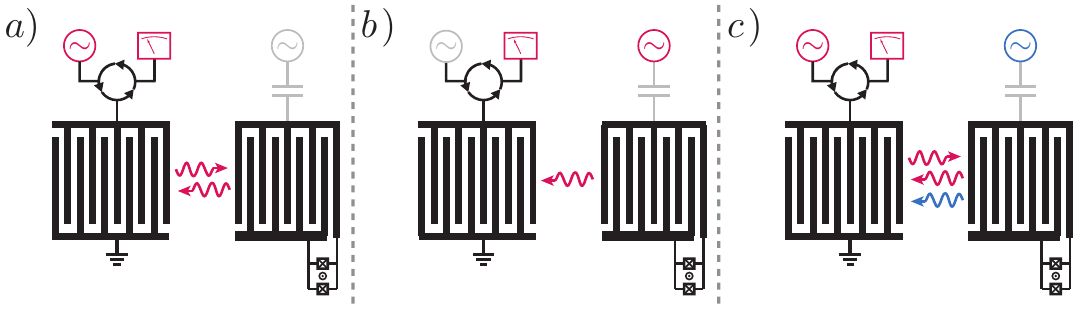}
\caption{ Three different experiments. (a) Acoustic reflection. In the first experiment an acoustic wave is launched towards the qubit, and the acoustic reflection is measured. (b) Listening. The qubit is excited through the gate (by a continuous RF signal or by an RF pulse), and the emission of phonons is detected by the IDT. (c) Two-tone spectroscopy. The acoustic reflection is measured while irradiating the qubit with microwaves through the gate.}
\label{fig:ThreeExp}
\end{center}
\end{figure}

1) On electrical and acoustical resonance, the reflection of SAW power from the qubit is nonlinear in the excitation power. For low powers, $P_{SAW} \ll \hbar \Gamma_{ac}$, the qubit reflects the incoming SAW perfectly. As the power increases and the $\ket{1}$ state of the qubit becomes more populated, the reflection coefficient decreases. For $P_{SAW} \gg \hbar \Gamma_{ac}$, the reflection coefficient tends to zero.

\begin{figure}[t!]
\begin{center}
\includegraphics[width=\linewidth]{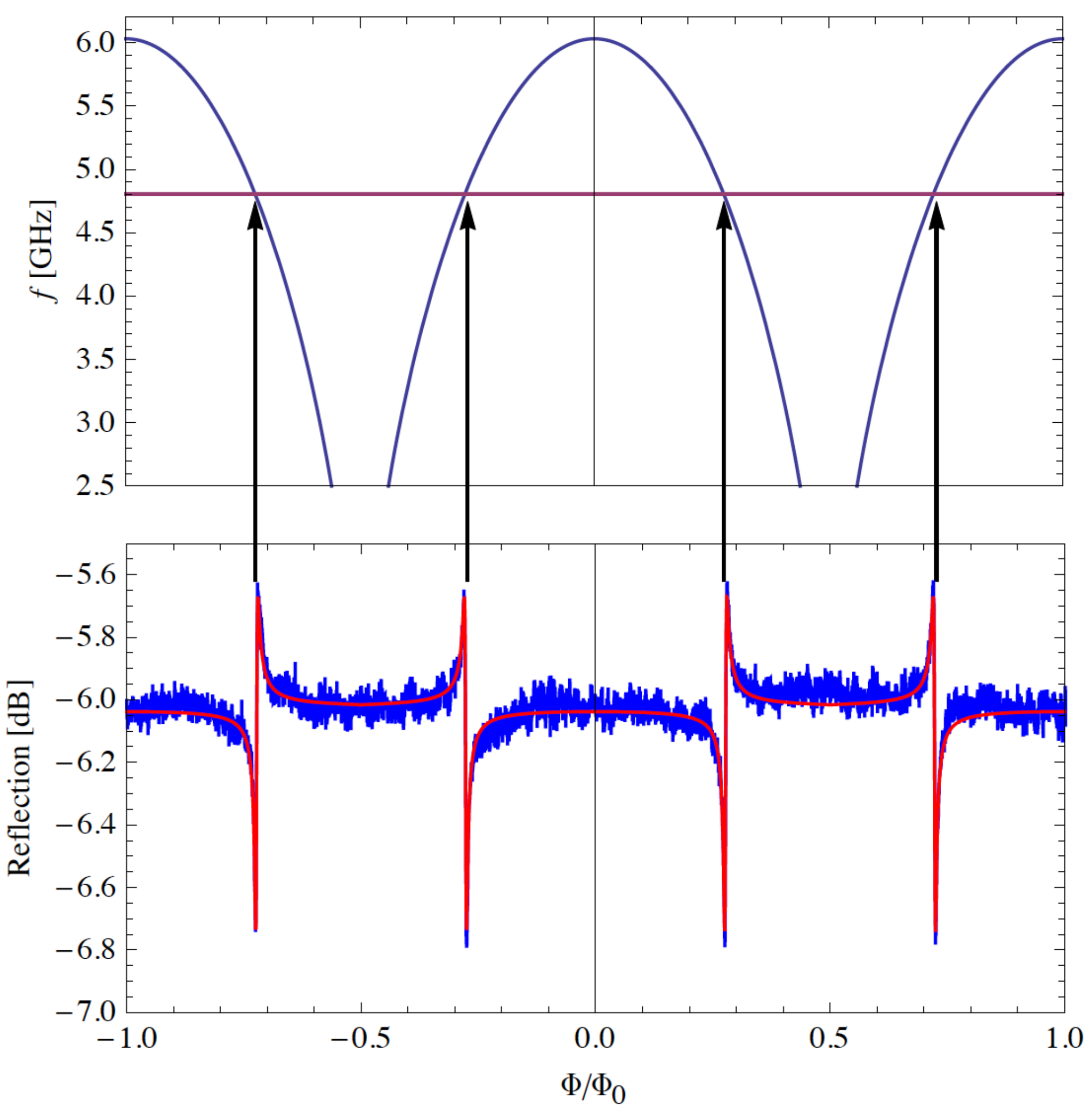}
\caption{(a) Qubit frequency as a function of external magnetic flux. The blue line is the calculated qubit frequency and the red line is the IDT frequency at which we can listen to the SAW. (b) Measured reflection coefficient from the IDT. In the flat regions the qubit is far detuned from the IDT frequency and the signal is just reflected from the IDT. When the qubit comes into resonance with the IDT frequency, there is an additional signal due to acoustic reflection at the qubit. The phase of the acoustic signal varies with the detuning and interferes with the signal which is directly reflected by the IDT. The blue trace is the measured data and the red trace is a fit to the theory.}
\label{fig:Reflection}
\end{center}
\end{figure}

2) The electrical resonance of the qubit can be tuned by applying a magnetic field through its SQUID loop, periodically bringing the electrical qubit resonance frequency $\omega_{10}$ in and out of the acoustic band of the IDT. This can be seen in \figref{fig:Reflection}.

3) When the qubit is excited through the gate at coinciding electrical and acoustic resonance frequencies, it relaxes by emitting SAW phonons, which can be detected by the IDT. The transmission from the gate to the acoustic channel is nonlinear in the same way as the acoustic reflection coefficient.

\begin{figure}[t!]
\begin{center}
\includegraphics[width=0.9\linewidth]{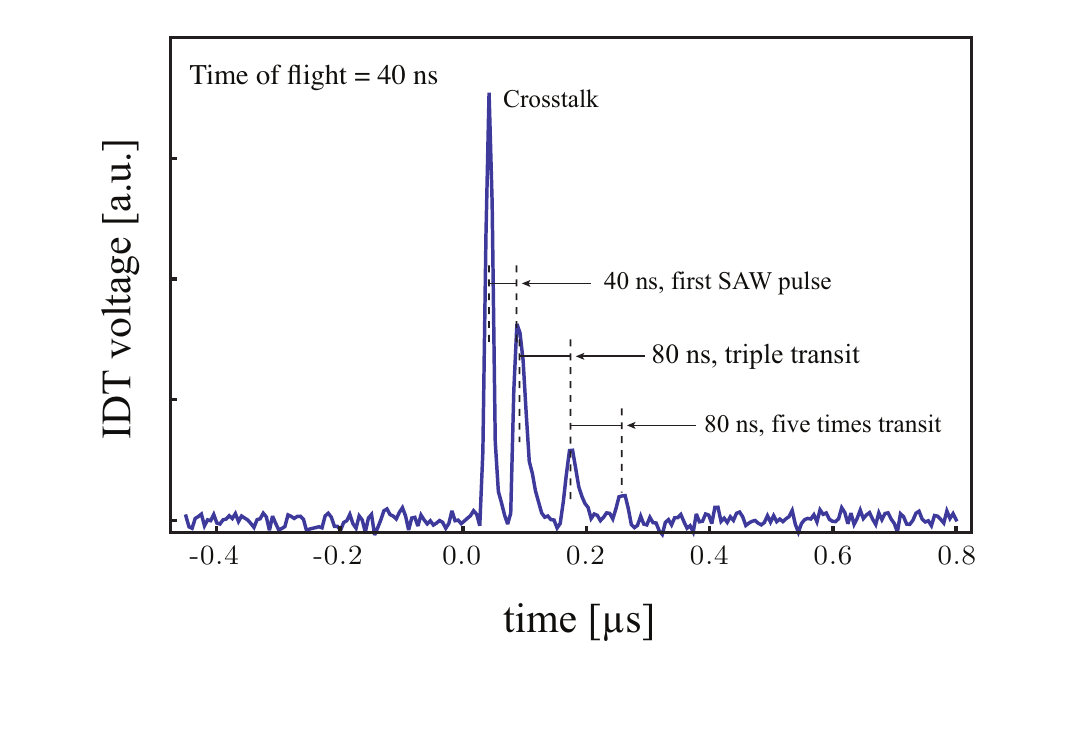}
\caption{Recorded IDT signal for short RF pulses applied through the gate. A $\unit[25]{ns}$ pulse is sent to the qubit gate at $t=0$. Due to a capacitive coupling between the qubit gate and the IDT, there is an immediate cross-talk response. The acoustic signal arrives after $\unit[40]{ns}$, which agrees well with the time of flight. Subsequent reflections of the acoustic signal give rise to additional echo signals spaced by $\unit[80]{ns}$.}
\label{fig:Echo}
\end{center}
\end{figure}

4) Since the electrical gate has a high bandwidth, it can be used to excite the qubit with short pulses at $\omega_{10}$. The emission from the qubit arrives at the IDT after a delay of $\sim \unit[40]{ns}$ compared with the applied electrical pulse. This corresponds to the acoustic propagation time between the qubit and the IDT.
In \figref{fig:Echo} we show how a $\unit[25]{ns}$ pulse is bouncing back and forth between the qubit and the IDT. The first peak is due to electrical cross-talk between the qubit gate and the IDT. The subsequent peak, which arrives $\unit[40]{ns}$ after the excitation pulse is applied, is the acoustic signal emitted by the qubit. The SAW is then partially reflected by the IDT and returns to the qubit, where it is reflected again. This echo signal arrives $\unit[80]{ns}$ after the pulse is applied. Two  echoes spaced by $\unit[80]{ns}$ can be observed. 

5) When the qubit is probed with a weak acoustic tone, its reflection coefficient depends in a complex way on the frequency and power of an electrical signal applied to the gate, as well as the electrical detuning of the qubit with respect to the acoustic center frequency. The features observed include an enhanced acoustic reflection at the $\ket{1}\rightarrow\ket{2}$ transition frequency when the $\ket{0} \rightarrow \ket{1}$ transition is addressed electrically, and Rabi splitting of the various energy levels when the electrical signal is strong.

All experiments show good agreement with the theory of \secref{sec:Theory}. The acoustic coupling rate also agrees with the semiclassical estimate given by \eqref{eq:classical_coupling}. Points 1 to 3 show that the qubit works as a two-level system, where the $| 0 \rangle \rightarrow | 1 \rangle$ transition can be addressed separately from all transitions to higher energy states. Point 4 proves directly that the qubit primarily relaxes by emitting SAW phonons. Point 5 underscores the nonclassical nature of the qubit, and demonstrates that the system is well described by the  quantum theory.

\section{Future directions}
\label{sec:Future}

As we have seen in the previous sections, quantum SAW devices have many similarities to circuit-QED devices. It is important to realize that there are also differences between the two. Although one may argue that photons will always be more coherent than phonons and therefore ask why one would be interested in SAW phonons at all, not all the differences are detrimental for experiments. On the contrary, below, we show that there are several interesting research directions to pursue and that the SAW phonons indeed offer new and interesting physics. The much lower propagation speed of SAWs compared to electromagnetic waves plays a crucial role here.

%It is important to realize that there are also interesting differences, but is there anything that makes the SAW phonons interesting when compared with photons?
%One may argue that photons will always be more coherent than phonons and therefore ask why one would be interested in SAW phonons at all.

\subsection{In-flight manipulation}
The speed of SAWs is of the order of $\unit[3000]{m/s}$ in most of the interesting piezoelectric materials, which is five orders of magnitude slower than for light in vacuum. This means, for instance, that the time it takes for a SAW signal to traverse a chip which is $\unit[3]{cm}$ long is approximately $\unit[10]{\mu s}$, while the corresponding traversal time for a light signal is just $\unit[300]{ps}$. There is ample time to manipulate a SAW signal ``in flight', but it would be very difficult to do something similar with photons, especially if one wants to do measurements and provide feedback signals in real time.
%If one wants to manipulate a SAW signal ``in flight'' there is plenty of time, but it would be very hard to do with photons, especially if one wants to do measurements and provide feedback signals in real time. 
Since the typical tuning time for a SQUID is less than $\unit[10]{ns}$ \cite{SandbergAPL2008,PierreAPL2014}, it would be possible to tune a transmon or a cavity in or out of resonance with the SAW wave many times during a $\unit[10]{\mu s}$ traversal.
\subsection{Coupling to optical photons}
Circuit QED systems need to be cold, both to maintain superconductivity and to avoid thermal excitations. On the other hand, it is clear that photons propagating either in free space or in optical fibers is the preferred choice for sending quantum information over large distances. The frequencies at which circuit QED devices work is five orders of magnitude lower than optical frequencies. This results in very similar wavelengths for SAW waves and optical signals. In the paper by Gustafsson \textit{et al.}~\cite{Gustafsson2014}, the wavelength used was $\unit[600]{nm}$. This wavelength could easily be modified to match with photons that could travel in optical fibers. Of course, the details of how such a conversion device would look like remain to be worked out. We note that many of the piezoelectric materials are transparent at optical frequencies and are routinely used in optical applications. It is also interesting that some of these materials have strong nonlinearities, especially $\textrm{LiNbO}_{3}$, which is used in optical modulators and other applications \cite{BoydBook}.
\subsection{Ultrastrong coupling between SAWs and artificial atoms}
In the SAW-qubit experiment, which is described above and in greater detail in Ref.~\cite{Gustafsson2014}, the coupling between the transmon and the SAW was $\unit[38]{MHz}$. In spite of the fact that GaAs is a very weakly piezoelectric material, this is similar to the coupling which has been shown between a transmon and open transmission lines in waveguide QED. Using other materials such as LiNbO$_3$ would allow much stronger coupling. Since the value of K2 is 70 times larger for LiNbO$_3$ than for GaAs (see \tabref{tab:MaterialProperties}), it should  be possible to reach couplings exceeding $\unit[1]{GHz}$, entering the ultrastrong coupling regime, or even the deep ultrastrong regime \cite{Niemczyk,Ballester}. In that situation, it is necessary to use a qubit which has a larger anharmonicity than the transmon, since the maximum anharmonicity of the transmon is about $\unit[10]{\%}$ of the qubit frequency. One possibility is to use a single-Cooper-pair box (SCB) \cite{Buttiker87,Bouchiat}, which has much larger anharmonicity. The fact that the SCB has a shorter coherence time is less of a problem since the important timescale is the inverse coupling rate, which is made very small in the ultrastrong coupling regime.
\subsection{Large atoms}
For experiments in quantum optics using natural atoms and laser light at visible wavelengths, the wavelength exceeds the size of the atom by several orders of magnitude. Rydberg atoms and superconducting qubits are much larger than regular atoms, but nonetheless substantially smaller than the wavelength of the radiation they interact with. In quantum acoustics with SAW waves, this small-atom approximation no longer holds. The transmon used in a SAW experiment can be of similar size to those used for circuit QED experiments, but the wavelength of the SAWs is orders of magnitude lower. Such devices thus operate in an unexplored regime, where the frequency dependence of the coupling strength and the Lamb shift are modified.  These modifications are discussed in \secref{sec:QuantumTheory} and in more detail in Ref.~\cite{KockumPRA2014}, both the coupling and the Lamb shift are modified. Further work could also explore the combination of large atoms with additional large atoms or other systems, and also the regime where the traveling time across the large atom is no longer negligible compared to the relaxation time of the atom.
\subsection{SAW resonators}
Initial experiments indicate that SAW resonators can reach quality factors of order $10^5$ at GHz frequencies, close to what is seen in superconducting transmission line resonators used in circuit QED \cite{Goppl2008}. Experiments with a transmon in an open SAW transmission line \cite{Gustafsson2014} indicate that coupling strengths can also be similar to those found in circuit QED, and higher still if strongly piezoelectric materials are used. These early results bode well for realizing strong coupling SAW-based circuit QED. Beyond such a basic realization, there are several experiments that can highlight differences between SAW resonators and conventional circuit QED.
%Beyond such a basic realisation, the special features of the SAW device may lead to new experimental regimes, and quantum information applications that are hard or impossible to reach with conventional circuit QED. 
For example, combining the in-flight manipulation discussed above with a SAW resonator could allow generation of exotic cavity phonon states, while the multimodal nature of long SAW cavities (similar to optical Fabry-Perot cavities) could enable new regimes of quantum optics to be reached. Advancing further the understanding of internal quality factors may push SAW resonators into a regime in which they may be useful as on-chip quantum memories, much smaller than their electromagnetic counterparts. Finally, it is worth noting that high-Q SAW resonators could also be employed to implement a new form of microwave cavity optomechanics \cite{SpringerCOMBook}, using circuit designs that realise a modulation of the qubit frequency by the SAW amplitude. Such an implementation may be of strong interest due to the combined high frequency and quality factor of the SAW compared to other mechanical systems used in the field.
\subsection{Analogues of quantum optics}
Finally, we note that there are a number of interesting quantum optics experiments which could be repeated in the quantum acoustics domain. For instance, a generator of single phonons should be possible to make in a similar way to single photon generators in circuit-QED. Engineering the couplings of a three level atom should allow the creation of population inversion and thus ``SAW-lasing''. 

\section{Conclusions}

In this paper, we have discussed new possibilities of performing quantum physics experiments using surface acoustic waves. We have showed that a superconducting transmon qubit can couple strongly to propagating SAWs and that the SAWs can be confined in high-Q resonators. Taken together, these results indicate that experiments conceived for quantum optics with photons can now be performed in quantum acoustics using SAW phonons. Furthermore, the slow propagation velocity and short wavelength of the phonons promises access to new regimes which are difficult to reach with traditional all-electrical circuits, such as giant atoms and improved feedback setups. %All in all, the foundation for quantum acoustics with SAWs has been laid and there are now many interesting directions to explore.

\section*{Acknowledgements}

This work was supported by the Swedish Research Council, the European Research Council, the Knut and Alice Wallenberg Foundation, the UK Engineering and Physical Sciences Research Council. We also acknowledge support from the People Programme (Marie Curie Actions) and the FET-project SCALEQIT of the European UnionÕs Seventh Framework Programme.

% BibTeX users please use
\bibliographystyle{apsrev}
\bibliography{SAWChapter}

\end{document}